\documentclass[a4paper,fleqn]{cas-sc}

\usepackage{amsmath}
\usepackage{amsthm}
\usepackage{mathtools}
\usepackage{tikz}
\usetikzlibrary{arrows.meta,calc}

\usepackage[ruled,vlined,noend]{algorithm2e}
\usepackage{quiver}
\ExplSyntaxOn
\cs_if_exist:cT { not= } { \cs_undefine:c { not= } }
\cs_if_exist:cT { not< } { \cs_undefine:c { not< } }
\cs_if_exist:cT { not> } { \cs_undefine:c { not> } }
\ExplSyntaxOff

\usepackage{fontspec}

\usepackage{unicode-math}
\usepackage{complexity}

\usepackage[nameinlink,capitalize]{cleveref}

\usepackage[numbers]{natbib}

\def\S{\ensuremath{\mathcal{S}}}

\def\D{\ensuremath{\mathcal{D}}}

\def\A{\ensuremath{\mathcal{A}}}
\def\B{\ensuremath{\mathcal{B}}}

\def\I{\ensuremath{\mathcal{I}}}
\def\O{\ensuremath{\mathcal{O}}}

\def\K{\ensuremath{\mathcal{K}}}

\newcommand{\ostar}[2]{\mathrm{St}^\circ(#1,#2)}

\newcommand{\chdiv}[1]{\mathrm{Ch}\ #1}
\newcommand{\smset}[0]{\{\mathsf{C},\mathsf{AS},\mathsf{IS}\}}
\newcommand{\fip}[1]{\mathsf{FI}_\mathsf{#1}}
\newcommand{\iter}[0]{\mathit{ITER}}

\newcommand{\ceil}[1]{\lceil {#1} \rceil}
\newcommand{\argmax}{\mathop{\mathrm{arg\,max}}\displaylimits}

\SetKwInput{KwShared}{Shared}
\SetKwInput{KwInitial}{Initial}

\def\REVIEW{0}  

\if \REVIEW 1
    \newcommand{\guillermo}[1]{{\color{blue}[Guillermo: #1]}}
    \newcommand{\petr}[1]{{\color{red}[Petr: #1]}}
\else
    \newcommand{\guillermo}[1]{}
    \newcommand{\petr}[1]{}
\fi

\newtheorem{theorem}{Theorem}[section]
\newtheorem{lemma}[theorem]{Lemma}
\newtheorem{proposition}[theorem]{Proposition}
\newtheorem{corollary}[theorem]{Corollary}

\theoremstyle{definition}
\newtheorem{definition}[theorem]{Definition}

\theoremstyle{remark}

\begin{document}
\let\WriteBookmarks\relax
\def\floatpagepagefraction{1}
\def\textpagefraction{.001}

\shorttitle{The Complexity of Full-Information Iterated Protocols}
\shortauthors{G. Toyos-Marfurt and P. Kuznetsov}

\title[mode=title]{The Complexity of Full-Information Iterated Protocols}

\tnotemark[1]
\tnotetext[1]{This article combines results that appeared, in a preliminary form, in~\cite{prequel1,prequel2}. The authors gratefully  acknowledge the support provided by the DUCAT project (grant ANR-20-CE48-0006).}


\author[1,2]{Guillermo Toyos-Marfurt}
\cormark[1]
\ead{guillermo.toyos@telecom-paris.fr}

\author[1]{Petr Kuznetsov}
\ead{petr.kuznetsov@telecom-paris.fr}


\affiliation[1]{organization={LTCI, T\'el\'ecom Paris, Institut Polytechnique de Paris},
city={Palaiseau},
country={France}}

\affiliation[2]{organization={SAMOVAR, T\'el\'ecom SudParis, Institut Polytechnique de Paris},
city={Palaiseau},
country={France}}

\cortext[1]{Corresponding author}


\begin{abstract}
Full-information protocols are a common abstraction in distributed computing. In its iterated form, processes proceed through successive communication rounds in which each process shares its entire current knowledge of the execution. 
%
%
Usually, it is implicitly assumed that a process can transmit an unbounded amount of information in every round.
This paper makes that assumption explicit and studies its computational cost. 

We formalize the problem of simulating iterated full-information protocols when communication is bounded, so that each process may share at most $b$ bits with every other process per round. 
%
%
We show that for a large family of shared-memory communication models and $n+1$ processes, simulating $r$ iterations of the full-information protocol requires
\(\Omega\!\left((n!)^{r-1} 2^{\,n-b}\right)\) rounds.
We complement this lower bound with simulation algorithms that are asymptotically optimal in the conventional read-write model.
%
In particular, the case of two processes is special: $r$ iterations of a full-information protocol can be simulated in exactly $r$ rounds while using only $2$-bit variables, as the amount of information that must be represented remains bounded for two processes but grows with each iteration for $n \geq 2$. 
\end{abstract}



\maketitle

\section{Introduction}
A fundamental difficulty in distributed computing is coordinating processes while handling partial and evolving views of the system arising from asynchrony, unreliable communication, and failures. 
In order to solve a distributed task, processes must exchange information so that they gradually gain knowledge about the states of the other processes.

\textit{Full-information protocols} provide a useful abstraction for reasoning about information exchange~\cite{HERLIHY20133,winkler2024time}.
In this abstraction, each process repeatedly communicates its entire current knowledge of the system to every process.
By assuming that processes always communicate all the information they possess, the model abstracts away the question of what is communicated and focuses solely on the effects of asynchronous executions.
Consequently, full-information protocols capture the maximal knowledge that processes can accumulate during an execution.

Many theoretical models formalize this idea through \emph{communication rounds}~\cite{mosesCons,HerSAS90,Raj10Shared,ELRAD1982155}. 
In such \emph{iterated} models, processes proceed through successive rounds in which they exchange information and update their local states based on the information received in that round. 
Iterated models provide a convenient framework for studying how knowledge accumulates over time and play a central role in theoretical analyses of distributed computation.

In particular, iterated full-information protocols underlie topological approaches to distributed computability~\cite{distCompTopo,HS99}. 
In this framework, the set of possible executions of a distributed model is represented as a simplicial complex, enabling combinatorial characterizations of several iterated shared-memory models~\cite{HS99,GKM14-podc,KRH18}. 
Within this setting, the celebrated Asynchronous Computability Theorem (ACT)~\cite{HS99} characterizes wait-free task computability for read-write shared-memory systems by relating solvability to the existence of chromatic maps between input and output complexes. 
The core observation behind ACT and its follow-ups~\cite{HR10,GKM14-podc,KRH18} is that finite executions of the read-write model can be described as iterations of the \emph{immediate snapshot}~\cite{BG93a,BG97,GR10-opodis}.
It turns out that the \emph{iterated immediate snapshot} model (IIS) has a very nice regular structure, precisely captured by iterative applications of the \emph{standard chromatic subdivision}~\cite{chromsubdivision} (Figure~\ref{fig:chromatic}).
This characterization relies on processes executing an iterated full-information protocol.

Full-information protocols, however, implicitly assume that processes can transmit an unbounded amount of information in every round. 
In practice, communication bandwidth and memory are bounded, and implementing a full-information protocol requires simulating full-information using messages of limited size. 
Interestingly, in wait-free read-write iterated models, unbounded memory is not necessary to preserve computational power: variables of minimal (two-bit) capacity suffice to implement arbitrarily many iterations of the full-information protocol, at the cost of additional rounds~\cite{delporte2024computational}. 
This observation raises a natural question: what is the cost of bounded communication?

In this paper, we study this question for a family of generic iterated algorithms, which we denote by $\mathit{ITER}$, that implement iterated full-information protocols. 
This family captures several standard communication models, including the Iterated Immediate Snapshot (IIS)~\cite{ImmediateSnap}, Iterated Atomic Snapshot (IAS)~\cite{AfekSnapshots93}, and Iterated Collect (IC)~\cite{AttiyaWelchBook} models. 
Our goal is to determine how many rounds an algorithm in $\mathit{ITER}$ must execute so that its protocol complex becomes isomorphic to the complex produced by $r$ iterations of the full-information protocol.

Our approach relies on the notion of process \emph{distinguishability} in the protocol complex --- a topological space representing the possible executions of an algorithm --- which we use to determine unique encodings of full-information states using a bounded number of bits. 
Concretely, for a fixed local state of a process, each state that some other processes may be in must be assigned a distinct encoding in order to remain distinguishable.
Thus, the number of encodings required is proportional to the number of such states, which corresponds precisely to the degree of vertices in the protocol complex.
To bound this quantity, we study how vertex degrees evolve under iterative applications of the standard chromatic subdivision.
This leads to a combinatorial problem of independent interest: enumerating the number of faces (i.e., joint states of process subsets) generated at each iteration.
We derive recursive equations that explicitly characterize the $f$-vector of the iterated chromatic subdivisions of any simplicial complex.
To the best of our knowledge, this is the first explicit characterization of the $f$-vector for this family of complexes.
This characterization is the key combinatorial ingredient enabling us to obtain asymptotic bounds on vertex degrees as a function of the number of processes $n$ and the number of subdivision iterations $r$.

\paragraph{Contributions.}
For a given input complex $\I$, we establish the trade-off between the number of communication rounds and the number of bits $b$ available per process per round needed to simulate $r$ iterations of a full-information protocol.
For systems with $n+1$ processes, we prove a lower bound of $\Omega( {(n!)^{r-1} \cdot 2^{n-b}})$ rounds.
We also present a simulation algorithm, called Greedy Star, that is asymptotically optimal in the iterated collect model, and runs in at most $O((n!)^{r-1} \cdot 2^{\,n-b} \cdot n)$ rounds in the atomic snapshot and immediate snapshot models.
In the particular case of simulating $r$ iterations in exactly $r$ rounds, we establish a tight bound of $\Theta(r n \log n)$ on the bit complexity $b$ for the collect and snapshot-based models.

To obtain these results, we derive novel recursive equations that characterize the $f$-vector of the iterated chromatic subdivisions of a simplicial complex --- that is, the vector which enumerates the faces in each dimension --- a classical tool from polyhedral combinatorics~\cite{fvector}.

Finally, we show that for two processes the round complexity is exactly $1$, achievable with only $2$-bit registers.
As a concrete illustration, we give an explicit construction of the full-information protocol for two processes using two-bit registers.
Using this construction, we obtain an implementation of approximate agreement~\cite{aproxag} in which the achievable precision corresponds to the number of simulated iterations of the full-information protocol.

%

\paragraph{Roadmap.}
The paper is organized as follows. 
We overview the related work in Section~\ref{sec:related}. 
In Section~\ref{sec:Preliminaries}, we recall the combinatorial topology tools and concepts used throughout this work.
In Section~\ref{sec:mymodel}, we describe the iterated full-information protocol and the family of iterated algorithms $\mathit{ITER}$.
Then in Section~\ref{sec:necessary}, we present the necessary conditions that an algorithm in $\mathit{ITER}$ must satisfy in order to have a protocol complex equivalent to the full-information protocol.
Next, in Section~\ref{sec:fvector}, we combinatorically characterize these conditions by studying the ${f}$-vector of the iterated chromatic subdivision. 
Finally, in Section~\ref{sec:results}, we present the main results on the asymptotic bounds for simulating the full-information protocols for algorithms in $\iter$.
In particular, Section~\ref{sec:greedy_star} introduces Greedy Star, a construction for obtaining bounded full-information protocols.
The case of a $2$-process system is studied in detail in Section~\ref{sec:2p}.
Finally, in Section~\ref{sec:end}, we conclude the paper and discuss open questions.

\section{Related Work}
\label{sec:related}

Wait-free immediate snapshot~\cite{BG93a} was introduced as a restriction of atomic snapshot memory~\cite{AfekSnapshots93}, where an execution can be seen as a series of blocks of updates followed by matching atomic snapshots.
In contrast, the collect model~\cite{attiyaCollect} is a relaxation where snapshots are replaced by sequences of atomic read operations.

A topological characterization of asynchronous computability is established by the Asynchronous Computability Theorem (ACT)~\cite{HS99}.
Among its consequences is the impossibility of wait-free set agreement, resolving a long-standing open problem~\cite{SZ93,BG93a}.
In this characterization, the information gained by processes across rounds is represented by successive standard chromatic subdivisions of the input complex, which correspond to the set of executions yielded by the Iterated Immediate Snapshot model (IIS).
A precise correspondence between executions of the IIS model and chromatic subdivisions is established and further studied in~\cite{GR10-opodis}.
It is formally proven that the IIS model corresponds to the iterated chromatic subdivision of the input complex~\cite{chromsubdivision}. 
A comprehensive treatment of distributed computing through combinatorial topology is given in~\cite{distCompTopo}.

Full-information protocols are also closely related to epistemic approaches to distributed computing, which model how processes accumulate and share knowledge during an execution. 
In this setting, information flow can be represented using subdivisions of simplicial complexes~\cite{gob21,gob26}.

Several works refine or extend the iterated immediate snapshot framework. 
Variants of the model have been proposed to study execution-time complexity~\cite{hoest} and to reduce the number of iterated layers required to solve tasks via simplex approximation~\cite{nishi}. 
More generally, equivalence relations between iterated and non-iterated adversarial read-write models have been established~\cite{bouzidStrg}, and task solvability has been characterized in broad classes of adversarial iterated models~\cite{KRH18}.

Another line of work studies the computational power of restricted communication. 
The wait-free solvability of tasks using $1$-bit messages in dynamic networks was analyzed in~\cite{2process_complexity}. 
More recently, the computational power of bounded registers in the wait-free and $t$-resilient models were characterized in~\cite{delporte2024computational}. 
%
%
Unlike those works, which focus on computability, this work studies the trade-off between round and bit complexity of simulating an iterated full-information protocol.

\section{Preliminaries}
\label{sec:Preliminaries}

\begin{figure}
    \centering
    \includegraphics[width=0.6\columnwidth]{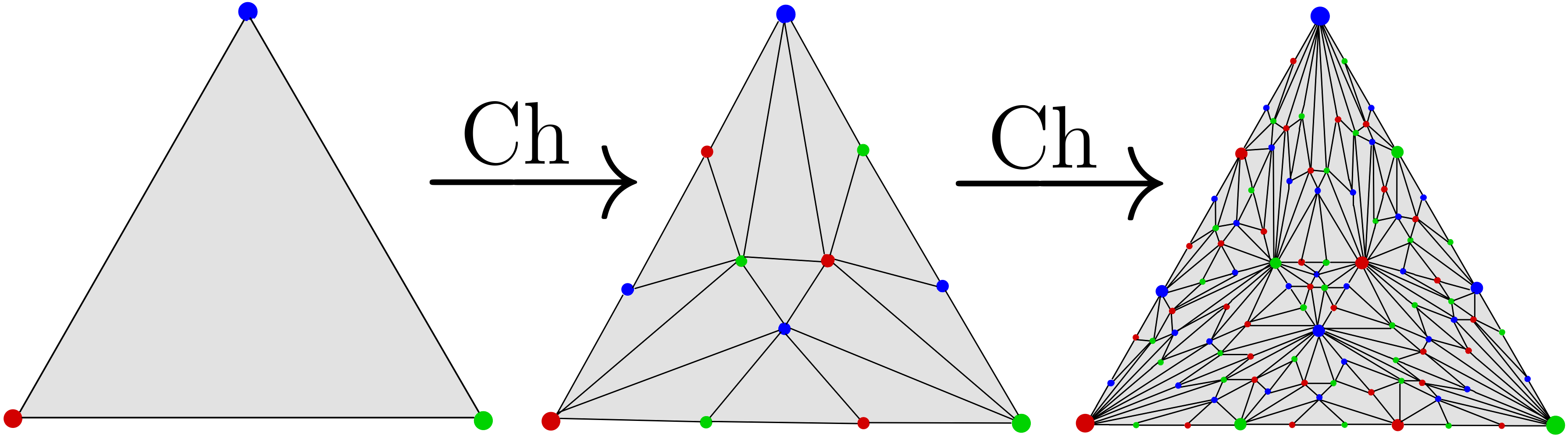}
    \caption{Iterative application of the standard chromatic subdivision to $\Delta_2$. Leftmost diagram illustrates the $\Delta_2$ simplex. The diagram in the middle shows $\mathrm{Ch}\ {\Delta_2}$, which is subdivided again resulting in the rightmost image $\mathrm{Ch}^2\ {\Delta_2}$.}
    \label{fig:chromatic}
\end{figure}

\paragraph{Simplicial complex.} 
We represent the states of iterated shared-memory protocols as topological-combinatorial spaces, called \emph{abstract simplicial complexes}~\cite{distCompTopo} --- i.e., a set of \emph{vertices} and an inclusion-closed set of vertex subsets, called \emph{simplices}. 
We call the simplicial complex of $n+1$ vertices with its power set the $n$-dimensional simplex $\Delta_n$.
The leftmost triangle in Figure~\ref{fig:chromatic} corresponds to $\Delta_2$.
The \emph{dimension} of a simplex $\Delta$, denoted as $dim(\Delta)$, is its number of vertices $V(\Delta)-1$.
The dimension of a simplicial complex is equal to the dimension of the largest simplex it contains.
A complex $\B$ is a subcomplex of $\A$ if every simplex of $\B$ is also a simplex of
$\A$. 
We call \textit{faces} the simplices contained in a simplicial complex, where a $0$-face is a vertex.
%
%
%
We call \textit{facets} the simplices which are not contained in any other simplex. 
A complex is \textit{pure} if all facets
have the same dimension. 
In this work, we only consider pure finite simplicial complexes.
We denote the set of vertices of a simplicial complex $\mathcal{A}$ as $V(\mathcal{A})$.
Two vertices are \emph{adjacent} if there is a $1$-face (an edge) in the simplicial complex containing both vertices.
We denote $\mathrm{deg}(\A,v)$ the number of vertices adjacent to $v$ in $\A$.
A simplicial complex is \textit{chromatic} if it is equipped with a \emph{coloring function} $\pi : V(\A) \rightarrow \Pi$ that assigns every vertex to a process identifier.
We assume that all simplicial complexes are chromatic.
Given a simplicial complex $\mathcal{A}$, the \emph{star} of $\mathcal{S}\subseteq\mathcal{A}$, $\mathrm{St}(\mathcal{A},\mathcal{S})$, is a subcomplex consisting of all simplices in $\mathcal{A}$ containing a simplex of $\mathcal{S}$ as face. 
We say two complexes $\A$ and $\B$ are isomorphic if there is a one-to-one color-preserving correspondence between their vertices that preserves the simplices --- that is, a set of vertices forms a simplex in $\A$ if and only if the corresponding set forms a simplex in $\B$. 
In other words, $\A$ and $\B$ have the same combinatorial structure, up to a relabeling of vertices.
Any abstract simplicial complex $\A$ admits a \textit{geometric realization}, denoted $|\A|$, constructed as follows:
we map its vertices to affinely independent points in $\mathbb{R}^{dim(\A)}$, and represent each simplex by the convex hull of its vertices. 
The resulting collection of simplices forms a \textit{geometric simplicial complex}, denoted $K(\A)$, and its union is its geometric realization $|\A|$.\guillermo{check.}

\paragraph{Carrier maps and subdivisions.}
A \emph{carrier map} $\Phi : \A \to 2^{\B}$ assigns simplices of $\A$ to subcomplexes of $\B$ such that for all $\tau,\tau' \in \A $ if $\tau' \subseteq \tau$ then $\Phi(\tau')\subseteq\Phi(\tau)$.
We say that $\Phi$ is \emph{strict} if for all $\tau,\tau' \in \A$, we have $\Phi(\tau \cap \tau') = \Phi(\tau) \cap \Phi(\tau')$.
We say that a complex $\B$ subdivides $\A$ if $|\A|=|\B|$ and each simplex of $K(\A)$ is the union of finitely many simplices of $K(\B)$.
A \textit{subdivision operator} is a strict carrier map $\tau : \mathcal{A} \to 2^{\mathcal{B}}$ with the following property: for any subcomplex $\mathcal{C} \subseteq \mathcal{A}$, the image $\tau(\mathcal{C})$ is a subdivision of $\mathcal{C}$.
%
%
Subdivisions are intersection preserving: for any subcomplexes $\A,\B \subseteq \I$ and subdivision operator $\tau$, $\tau(\A) \cap \tau(\B) = \tau(\A \cap \B)$.
The \emph{standard chromatic subdivision} of a simplicial complex $\mathcal{A}$, denoted as $\mathrm{Ch}\ \mathcal{A}$, is a complex whose vertices are tuples $(c,\sigma)$ where $c$ is a color and $\sigma$ is a face of $\mathcal{A}$ containing a vertex of color $c$. 
A set of vertices of $\mathrm{Ch}\ \mathcal{A}$ defines a simplex if for each pair $(c,\sigma)$ and $(c',\sigma')$, $c\neq c'$ and either $\sigma \subseteq \sigma'$ or $\sigma' \subseteq \sigma$. 
For $\Delta_n$, the vertices $(c,\Delta_n)$ define a \emph{central simplex} in $\mathrm{Ch}\ \mathcal{A}$, these vertices are called \emph{central vertices}.
The standard chromatic subdivision can be seen as the ``colored'' analog of the \emph{standard barycentric subdivision}~\cite{distCompTopo}. 
Figure~\ref{fig:chromatic} illustrates the application of $\mathrm{Ch}$ to the $2$-dimensional simplex.
Note that $\mathrm{Ch}$ is itself a subdivision operator.

\paragraph{Tasks.} 
A \emph{distributed task} is defined as a tuple $(\I,\O, \Delta)$, where $\I$ and $\O$ are, resp., the \emph{input complex} and the \emph{output complex}, describing the task's input and output configurations, and $\Delta: \I \rightarrow 2^{\O}$ is a map relating each input assignment to all output assignments allowed by the task.
In this work we consider general \emph{colored} tasks, where $\I$ and $\O$ are chromatic simplicial complexes, allowing processes to have different sets of inputs and outputs.
To solve a task, a distributed system uses a (communication) \emph{protocol} (i.e., an algorithm) to share information between processes.
The set of configurations reachable by the protocol from a given initial configuration  are expressed via a carrier map, denoted by $\Xi$.
%
%
We refer to this function as the \emph{protocol map}.
From the \emph{protocol complex} $\Xi(\mathcal{I})$, a 
simplicial map (i.e. a function that maps simplices to simplices)  determines the outputs of processes by mapping the simplices of the protocol complex to an output complex, respecting the specification of the problem given by $\Delta$: $\forall \sigma \in \I, \delta \circ \Xi(\sigma) \subseteq \Delta(\sigma)$. We refer to $\delta$ as the \emph{decision function}.
Note that, unlike linearizability~\cite{HW90}, correctness of a distributed task is expressed topologically: a protocol is correct if it maps every input simplex to an allowed output simplex respecting the task specification $\Delta$.
Let $\Xi$ and $\Xi'$ be two protocol maps. We say $\Xi$ and $\Xi'$ are \textit{equivalent} if $\mathrm{dom}(\Xi)=\mathrm{dom}(\Xi')$ and for every input complex $\I \in \mathrm{dom}(\Xi), \Xi(\I) \cong \Xi'(\I)$. That is, two protocol maps are equivalent if they always yield an isomorphic protocol complex.  

\section{System Model}
\label{sec:mymodel}

\subsection{Computational Model}
\label{sec:compmodel}
We consider a system of $\Pi=\{p_1,\ldots,p_{n+1}\}$ asynchronous processes, which may fail by crashing; a crashed process takes no further steps.
There are no timing assumptions and no bounds on the number of failures.
All protocols considered are \emph{wait-free}: each process reaches an output in a finite number of its own steps, regardless of the other processes.

\subsubsection{Communication Primitives}
Processes communicate via a sequence of \textit{communication primitives} $\mathsf{Comm}[1],\ldots,\mathsf{Comm}[r]$, for some fixed $r \geq 1$.
Each primitive $\mathsf{Comm}[k]$ supports two one-shot operations, \emph{publish} and \emph{collect}, and processes access the primitives in sequential order.
Let $I$ be a set of values.
Primitive $\mathsf{Comm}[k]$ allows each process $p_i \in \Pi$ to invoke $\mathsf{Comm}[k].\mathsf{publish}(i,s)$ with $s \in I$ at most once,
followed by at most one invocation of $\mathsf{Comm}[k].\mathsf{collect}()$, which returns a value $V_i^k \in \{\,V : \Pi \rightharpoonup I\,\}$.
We call the returned value $V_i^k$ the \emph{view} of process $p_i$ in round $k$.
We denote by $\mathrm{dom}(V_i^k) \subseteq \Pi$ the domain of $V_i^k$, that is, the set of processes whose values appear in the view.
The communication primitives satisfy the following properties.
For every process $p_i \in \Pi$ and round $1 \le k \le r$ such that $p_i$ completes $\mathsf{collect}$ returning $V_i^k$:
\begin{itemize}
\item\textbf{Self-inclusion.}
$p_i \in \mathrm{dom}(V_i^k)$.
\item\textbf{No fabrication.}
If $p_j \in \mathrm{dom}(V_i^k)$, then process $p_j$ invoked
$\mathsf{Comm}[k].\mathsf{publish}(j,s_j)$
before $p_i$ returned from $\mathsf{Comm}[k].\mathsf{collect}()$, and $V_i^k(p_j)=s_j$.
\item\textbf{Collect (RW Atomicity).} For all $p_j \in \Pi$ such that $p_j$ completes $\mathsf{collect}$ returning $V_j^k$, either $p_i \in \mathrm{dom}(V_j^k)$ or $p_j \in \mathrm{dom}(V_i^k)$.
\end{itemize}
The Collect predicate captures the behavior of atomic read/write shared memory, where each process writes an input and then reads the registers of other processes~\cite{AttiyaWelchBook}.
We denote this primitive by $\mathsf{Comm}_{\mathsf{C}}$.
Stronger shared-memory communication primitives are obtained by imposing additional structural constraints on views.
\begin{itemize}
    \item\textbf{Containment (Atomic Snapshot Object).} $\mathrm{dom}(V_i^k) \subseteq \mathrm{dom}(V_j^k)$ or $\mathrm{dom}(V_j^k) \subseteq \mathrm{dom}(V_i^k)$.
    Thus, the domains of views are totally ordered by inclusion.
    This corresponds to atomic snapshot objects~\cite{AfekSnapshots93}.
    We denote this primitive by $\mathsf{Comm}_{\mathsf{AS}}$.

    \item\textbf{Immediacy (Immediate Snapshot Object).} In addition to Containment, the
    following immediacy condition holds: $p_i \in \mathrm{dom}(V_j^k) \;\Rightarrow\; \mathrm{dom}(V_i^k) \subseteq \mathrm{dom}(V_j^k)$.
    This corresponds to immediate snapshot objects~\cite{ImmediateSnap}.
    We denote this primitive by $\mathsf{Comm}_{\mathsf{IS}}$.
\end{itemize}

For $\mathsf{A} \in \{\mathsf{C},\mathsf{AS},\mathsf{IS}\}$, let $\mathcal{V}_{\mathsf{A}}$ denote the set of all families of views $\{V_i^k\}_{p_i \in \Pi'\subseteq \Pi}$ that may be returned by completed invocations of $\mathsf{Comm}_{\mathsf{A}}.\mathsf{collect}()$.
\begin{lemma}[Hierarchy of communication abstractions] \label{lemma:viewh}
For $|\Pi|\geq 2$, the sets of families of views satisfy
$
\mathcal{V}_{\mathsf{IS}}
\subseteq
\mathcal{V}_{\mathsf{AS}}
\subseteq
\mathcal{V}_{\mathsf{C}}$, and these inclusions are strict when $|\Pi|\geq 3$.
\end{lemma}
\begin{proof}
Let $\{V_i\}_{p_i \in \Pi'\subseteq\Pi}$ be a family of views returned by $\mathsf{Comm}_{\mathsf{A}}.\mathsf{collect}()$.
Write $D_i := \mathrm{dom}(V_i)$.
\noindent\emph{Inclusions.}
Immediate Snapshot includes containment, hence $\mathcal{V}_{\mathsf{IS}} \subseteq \mathcal{V}_{\mathsf{AS}}$.
If containment holds, then for any $i,j$ either $D_i \subseteq D_j$ or $D_j \subseteq D_i$; by self-inclusion this implies $p_i \in D_j$ or $p_j \in D_i$, hence $\mathcal{V}_{\mathsf{AS}} \subseteq \mathcal{V}_{\mathsf{C}}$.
\noindent\emph{Strictness.}
To show $\mathcal{V}_{\mathsf{IS}} \subset \mathcal{V}_{\mathsf{AS}}$, consider $\Pi'=\{p_1,p_2,p_3\}$ with $D_2=\{p_1,p_2\}, D_1=\{p_1,p_2,p_3\}, D_3=\{p_1,p_2,p_3\}$.
The family satisfies containment since $D_2 \subseteq D_1 = D_3$, hence it lies in $\mathcal{V}_{\mathsf{AS}}$.
However, $p_2 \in D_1$ while $D_2 \nsubseteq D_1$ is false.
Thus the immediacy implication fails, and the family is not in $\mathcal{V}_{\mathsf{IS}}$.
To show $\mathcal{V}_{\mathsf{AS}} \subset \mathcal{V}_{\mathsf{C}}$, consider $\Pi'=\{p_1,p_2,p_3\}$ with $D_1=\{p_1,p_2\}, D_2=\{p_2,p_3\}, D_3=\{p_1,p_2,p_3\}$.
Every pair satisfies Collect (RW atomicity). Thus the family lies in $\mathcal{V}_{\mathsf{C}}$. Containment fails since neither $D_1 \subseteq D_2$ nor $D_2 \subseteq D_1$, hence the family is not in $\mathcal{V}_{\mathsf{AS}}$.
\end{proof}

\begin{lemma} \label{lemma:viewh2}
For $|\Pi|=2$, the sets of families of views satisfy
$\mathcal{V}_{\mathsf{IS}} = \mathcal{V}_{\mathsf{AS}} = \mathcal{V}_{\mathsf{C}}$.
\end{lemma}

\begin{proof}
Let $\Pi=\{p_1,p_2\}$ and consider any family of views $\{V_i\}_{i \in \Pi’} \in \mathcal{V}_{\mathsf{C}}$, with  $D_i =\mathrm{dom}(V_i)$.
By self-inclusion and the Collect property, at least one view is $\{p_1,p_2\}$, and the other is either the same or a singleton containing only the process itself.
In all cases, the domains are comparable by inclusion, so Containment holds, and Immediacy is automatically satisfied.
Hence every family in $\mathcal{V}_{\mathsf{C}}$ also lies in $\mathcal{V}_{\mathsf{AS}}$ and $\mathcal{V}_{\mathsf{IS}}$. Combined with the inclusions in Lemma~\ref{lemma:viewh}, this gives $\mathcal{V}_{\mathsf{IS}} = \mathcal{V}_{\mathsf{AS}} = \mathcal{V}_{\mathsf{C}}$.
\end{proof}

\subsubsection{Iterated Full-Information Protocols}
We now define the iterated full-information protocol executed over the sequence of communication primitives $\mathsf{Comm}[1],\ldots,\mathsf{Comm}[r]$.
The pseudocode executed by each process is given in Algorithm~\ref{alg:FIP}.
Each process starts with an initial input value drawn from an input domain $I$.
%
%
Let $\S^0 = I$ be the initial input state.
For each $k \geq 1$, define $\S^k = \{V: \Pi \rightharpoonup S^{k-1}\}$.
We call an element of $\S^k$ a \emph{$k$-layer view}.
Thus, a $k$-layer view is a partial function mapping processes to $(k-1)$-layer views.

The protocol proceeds in rounds. At the beginning of round $k$, the local state of process $p_i$ is some value $s_i^{k-1} \in \S^{k-1}$.
Process $p_i$ invokes $ \mathsf{Comm}[k].\mathsf{publish}(i, s_i^{k-1})$, and then invokes $\mathsf{Comm}[k].\mathsf{collect}()$, which returns a view $V_i^k \in \S^k$.
The process then updates its local state to $s_i^k := V_i^k$.
Thus, at each round, a process publishes its entire current state and replaces it with the view returned by the communication primitive.
After completing round $r$, the local state of process $p_i$ is a value in $\S^r$, representing its full-information view of the execution.
For $\mathsf{A} \in \{\mathsf{C},\mathsf{AS},\mathsf{IS}\}$, we denote by $\mathsf{FI}_{\mathsf{A}}$ the iterated full-information protocol executed over the sequence $\mathsf{Comm}_{\mathsf{A}}[1],\ldots,\mathsf{Comm}_{\mathsf{A}}[r]$.

\begin{algorithm}
\SetAlgoLined
\KwShared{$\textsf{Comm}[1,\ldots,r]$ \hfill {\textbf{\!/\!/} Sequence of round communication primitives.}}

\KwInitial{$s = \textsf{input}(i)$ \hfill {\textbf{\!/\!/} Initial input of process $p_i$.}}
\For{$k:=1$ to $r$}{
  $\textsf{Comm}[k].\textsf{publish}(i,s)$ \\
  $s$ \gets $\textsf{Comm}[k].\textsf{collect}()$ \\
}
\Return $s$\;
\caption{The iterated full-information protocol. Code for process $p_i \in \Pi$ and $r \geq 1 $ rounds.}
\label{alg:FIP}
\end{algorithm}

We now introduce \(\iter\), a family of iterated protocols which generalizes Algorithm~\ref{alg:FIP}.
$\iter$ captures all round-based protocols.
While in the full-information protocol processes share their entire local state in every round, protocols in $\iter$ may publish an arbitrary value at each round.
As in Algorithm~\ref{alg:FIP}, processes proceed in rounds using a sequence of communication primitives. 
However, instead of publishing their full state, processes apply an \emph{encoding function} before writing to the communication object. 
After collecting a view of the encoded values written by other processes, they update their local state according to a \emph{state transition function}.

Formally, we define the set of possible states inductively by round.
Let $I$ denote the input domain. The initial state space is $S^0 = I$.
For each round $k \leq r$ processes publish an encoded representation of their current state. Let $E_k$ denote the set of possible encoded values at round $k$.
The encoding function of round k is
$\omega_k : \Pi \times S^{k-1} \rightarrow E_k$, which specifies the value written by process $p_i$ when its current state is $s \in S^{k-1}$.
The $k$-level view returned by the collect operation consists of the encodings written by the processes that are observed in that round. 
Thus, the set of possible views at round $k$ is, $V_k = \{\, V : \Pi \rightharpoonup E_k \,\}$.
The process then updates its local state according to the state transition function $\sigma_k : \Pi \times S^{k-1} \times V_k \rightarrow S^{k}$.
Algorithm~\ref{alg:encodedFIP} describes the pseudocode executed by each process. 
For clarity, we denote $\sigma_k$ as $next\_state(\cdot,\cdot,k)$ and $\omega_k$ as $encode(\cdot,k)$.

\begin{algorithm}
\SetAlgoLined
\KwShared{$\textsf{Comm}[1,\ldots,r]$ \hfill {\textbf{\!/\!/} Sequence of round communication primitives.}}
\KwInitial{$v = \textsf{input}(i)$ \hfill {\textbf{\!/\!/} Initial input of process $p_i$.}}
\KwInitial{$s = next\_state(v,\bot,0)$ \hfill \textbf{\!/\!/} Initial state of process $p_i$.}
\For{$k:=1$ to $r$}{
  $\textsf{Comm}[k].\textsf{publish}(i,encode(s,k))$ \\
  $v$ \gets $\textsf{Comm}[k].\textsf{collect}()$ \\
  $s \gets next\_state(s,v,k)$ \\
}
\Return $s$\;
\caption{The $\iter$ family of protocols. Code for process $p_i \in \Pi$ and $r \geq 1 $ rounds.}
\label{alg:encodedFIP}
\end{algorithm}

The information each process sends at each round $k$ is determined by the encoding function $\omega_k$.
We say that an encoding function $\omega_k$ is \emph{bounded} if there exists a constant $B$ such that $|E_k| \le B$. In that case, each process shares at most $\lceil \log_2 (B+1) \rceil$ bits per round.
The bit complexity of an $r$-round protocol in $\iter$ is $\sum_{k=1}^{r} \lceil \log_2 |E_k + 1| \rceil$.

Observe that the full-information protocols described in Algorithm~\ref{alg:FIP} are instances of $\iter$. If for all $k\leq r$, $\omega_k(i,s) = s$ and $\sigma_k(i,s,v) = v$, then processes publish their entire state and replace it with the collected view.
In this case the encoding set of round k is $E_k = \S^{k-1}$, and the size of the state space grows exponentially with the number of rounds.

\begin{lemma}
The bit complexity of the full-information protocols of Algorithm~\ref{alg:FIP} is $\Theta(|\Pi|^{r-1})$.
\end{lemma}
\begin{proof}
Fix $k \leq r$.
At round $k=1$, each process publishes its input value. Since the input domain $\I$ is finite, this requires $\Theta(1)$ bits.
At round $k \geq 2$, each process publishes its entire $(k-1)$-layer view. Such a view contains at most one entry per process, and each entry stores a $(k-2)$-layer view. Thus the size of a $(k-1)$-layer view satisfies the recurrence $b_{k-1} = \Theta(|\Pi| \cdot b_{k-2})$,
where $b_k$ denotes the number of bits needed to represent a k-layer view.
Since $b_0 = \Theta(1)$, a straightforward induction gives
$b_k = \Theta(|\Pi|^k)$.
Therefore, the bit complexity of the protocol is $\sum_{k=1}^{r} \Theta(|\Pi|^{k-1}) = \Theta(|\Pi|^{r-1})$.
\end{proof}

In this work we study protocols in $\iter$ that are \textit{equivalent} to $r$ iterations of the full-information protocol $\fip{A}$, where $\mathsf{A} \in \{\mathsf{C},\mathsf{AS},\mathsf{IS}\}$ (i.e., their respective protocol maps are equivalent).
%
%
Our objective is to characterize the protocols in $\iter$ that simulate $r$ iterations of $\fip{A}$ while using \textit{bounded encodings}. 
For this purpose, we impose an explicit communication constraint on protocols in $\iter$: in every round each process may transmit at most $b$ bits to the other processes.
Hence, there are at most $2^b$ distinct messages that can be sent in a given round. 
In our model, we reserve one of these encodings to represent the absence of a message (denoted by $\bot$).
Therefore, the set of non-$\bot$ encodings satisfies $|E_k| \le 2^b-1$ for all rounds $k \le r$. 
Under this restriction, the encoding function $\omega_k$ is bounded and must share the information contained in the current local state into a representation of at most $b$ bits.

This restriction introduces a fundamental trade-off between the number of rounds and the quantity of information communicated per round.
In the unbounded full-information protocol, processes share their entire state in every round, which allows the protocol to simulate $r$ iterations of full-information in exactly $r$ rounds.
In contrast, protocols in $\iter$ with bounded encodings may require additional rounds to progressively reveal the information that would otherwise be transmitted in a single step of the full-information protocol.
The objective of this work is to characterize the number of rounds required for a protocol in $\iter$ with $b$-bit encodings to simulate $r$ iterations of the full-information protocol. 
This leads to a \textit{round–communication trade-off}: smaller encodings reduce the per-round communication cost but may increase the number of rounds needed to reproduce the knowledge structure of the full-information execution.

\subsection{Topological Model}
\label{sec:topomodel}

We denote by $\Xi_A$ the protocol map of an algorithm $A \in \mathit{ITER}$. 
Since $A$ uses a different pair of functions $(\omega_k, \sigma_k)$ at each round $k$, it induces a distinct protocol map $\Xi^*_{A(k)}$ for each round $k$.
We write $\Xi_{A(k)}(\I)$ to refer to the protocol complex obtained after applying the sequence of intermediate protocol maps from round $1$ to $k$: $\Xi_{A(k)}(\I) = \Xi^*_{A(k)} \circ \Xi^*_{A(k-1)} \circ \dots \circ \Xi^*_{A(1)}(\I)$.
Accordingly, we have that $\Xi_A = \Xi_{A(R)}$, were $R$ is the last round of $A$ (as defined in Algorithm~\ref{alg:encodedFIP}).

\begin{figure}
    \centering
    \includegraphics[width=0.6\columnwidth]{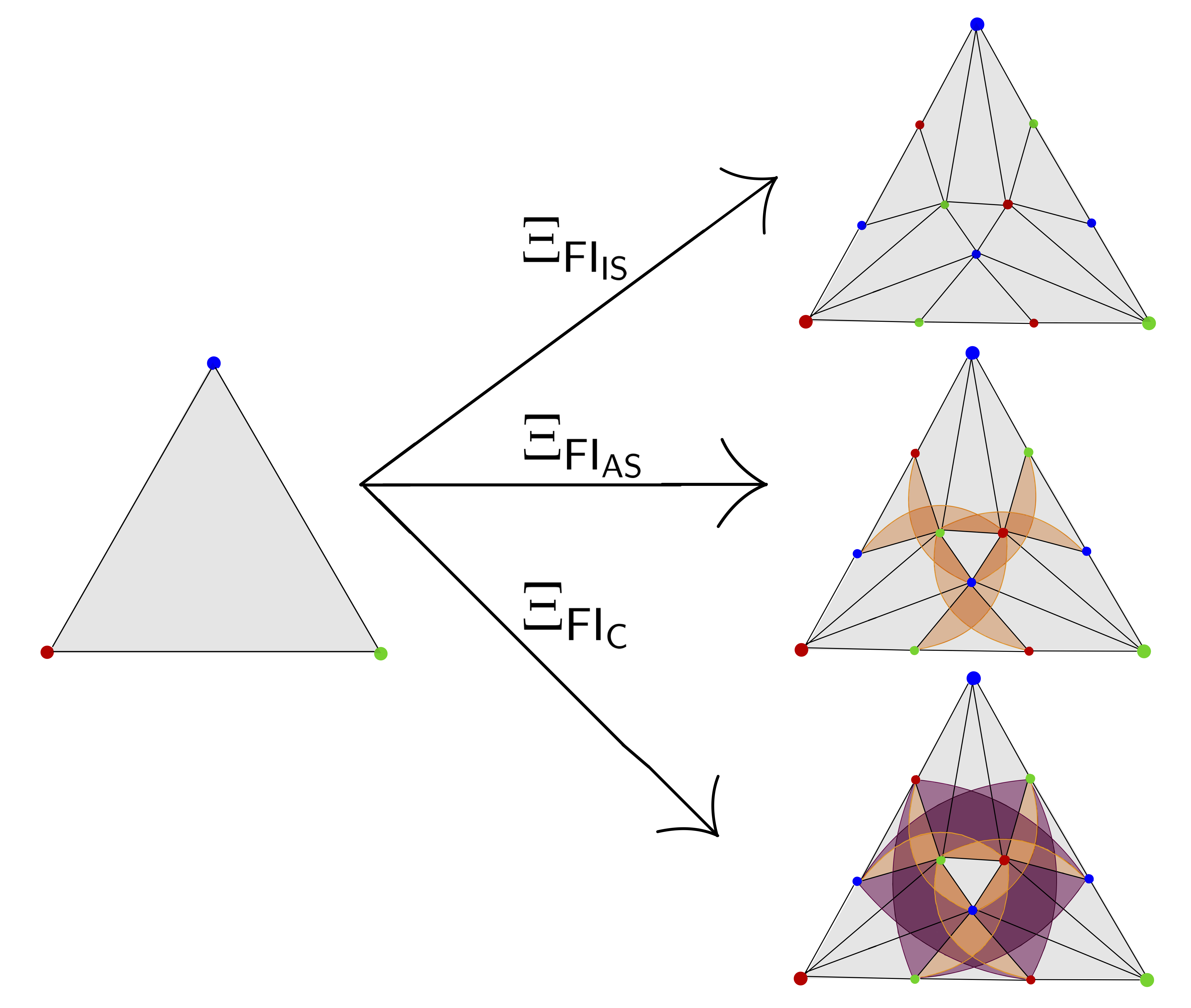}
    \caption{Application of the  \textit{immediate snapshot}, \textit{atomic snapshot} and \textit{iterated collect} full-information protocol maps to $\Delta_2$ (shown on the right, from top to bottom). 
    Starting from $\Xi_{\mathsf{FI}_\mathsf{IS}}(\Delta_2)$,
    the new possible configurations in the resulting protocol complexes of $\Xi_{\mathsf{FI}_\mathsf{AS}}$ and $\Xi_{\mathsf{FI}_\mathsf{C}}$ are highlighted in orange and purple, respectively. Note that $\Xi_{\mathsf{FI}_\mathsf{IS}}(\Delta_2) \subset \Xi_{\mathsf{FI}_\mathsf{AS}}(\Delta_2) \subset \Xi_{\mathsf{FI}_\mathsf{C}}(\Delta_2)$ and $\Xi_{\mathsf{FI}_\mathsf{IS}} \cong \mathrm{Ch}$.}
    \label{fig:subdivisions}
\end{figure}

Figure~\ref{fig:subdivisions} highlights the resulting protocol maps of one iteration of the full-information protocols in Algorithm~\ref{alg:FIP}.
We denote $\Xi_{\fip{A}}^r$ the resulting protocol map of $r$ iterations of the full-information protocol using a communication primitive $\mathsf{A} \in \smset$.
We look for bounded memory protocol maps that yield the same output complex as the unbounded full-information protocols.
That is, given a simplicial input complex $\mathcal{I}$ and an algorithm $A \in \mathit{ITER}$ using a communication primitive $\mathsf{A}$, we want the protocol complex $\Xi_A(\mathcal{I})$ to be \emph{isomorphic} to the protocol complex yielded by $\fip{A}$. Particularly, if $A$ uses immediate snapshot, we want that $\Xi_A(\mathcal{I}) \cong \Xi_{\fip{A}}(\mathcal{I}) \cong \chdiv{\I}$~\cite{chromsubdivision}.

\begin{figure}
\centering
\[\begin{tikzcd}
	{\Xi_{A(1)}(\mathcal{I})} & \dots & {\Xi_{A(k_1)}(\mathcal{I})} & \dots & {\Xi_{A(k_2)}(\mathcal{I})} & \dots & {\Xi_{A(k_r)}(\mathcal{I})} \\
	\\
	{\mathcal{I}} && {\Xi_{\mathsf{FI}_{\mathsf{A}}} ( \mathcal{I})} && {\Xi_{\mathsf{FI}_{\mathsf{A}}}^2 ( \mathcal{I})} & \dots & {\Xi_{\mathsf{FI}_{\mathsf{A}}}^r(\mathcal{I})}
	\arrow[from=1-1, to=1-2]
	\arrow[from=1-2, to=1-3]
	\arrow[from=1-3, to=1-4]
	\arrow[from=1-4, to=1-5]
	\arrow[from=1-5, to=1-6]
	\arrow[dashed, tail reversed, from=1-5, to=3-5]
	\arrow[from=1-6, to=1-7]
	\arrow[from=3-1, to=1-1]
	\arrow["{\Xi_{\mathsf{FI}_{\mathsf{A}}}}"', from=3-1, to=3-3]
	\arrow[dashed, tail reversed, from=3-3, to=1-3]
	\arrow["{\Xi_{\mathsf{FI}_{\mathsf{A}}}}"', from=3-3, to=3-5]
	\arrow["{\Xi_{\mathsf{FI}_{\mathsf{A}}}}"', from=3-5, to=3-6]
	\arrow["{\Xi_{\mathsf{FI}_{\mathsf{A}}}}"', from=3-6, to=3-7]
	\arrow[dashed, tail reversed, from=3-7, to=1-7]
\end{tikzcd}\]
\caption{Commutative diagram illustrating the equivalence between the unbounded full-information protocol map $\Xi_{\fip{A}}$ and the iterated protocol map $\Xi_A$, with $\mathsf{A} \in \smset$, over an input complex $\mathcal{I}$. $\Xi_A(k)$ denotes the associated protocol map at the k-th iteration of algorithm $A$. Dashed arrows indicate the existence of an isomorphism. The bounded algorithm $A$ is able to solve the same set of tasks as the unbounded full information protocol.}
\label{fig:iso-iter-diag}
\end{figure}
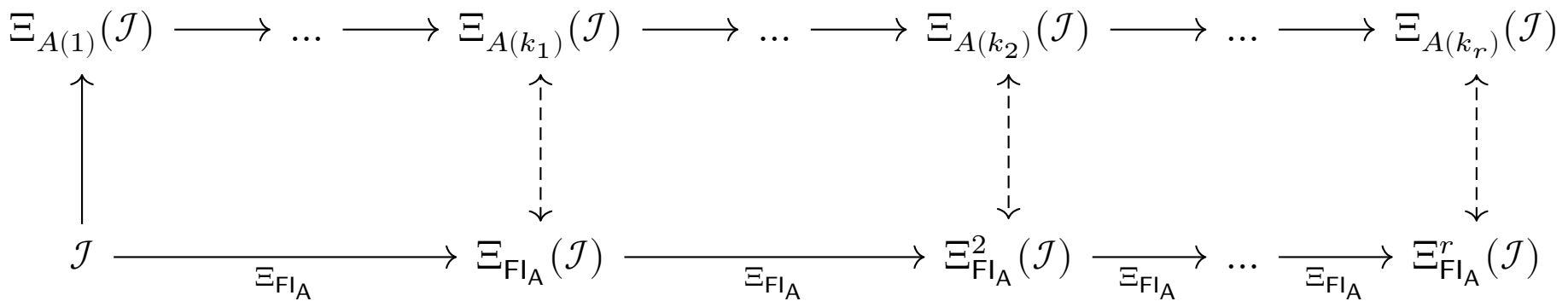

For each $k\leq r$, an algorithm $A \in \iter$ may execute an arbitrary number of rounds in order to obtain an equivalent output complex to $k$ iterations of the full-information protocol. 
Figure~\ref{fig:iso-iter-diag} presents a commutative diagram that illustrates this concept. 
For each $k\leq r$, there exists a round $t_k$ of algorithm $A$ such that the protocol complex of $A$ at round $t_k$ (i.e., $\Xi_{A(t_k)}$) is isomorphic to the protocol complex of $k$ iterations of the full-information protocol:  $\Xi_{A(t_k)} \cong \Xi_{\fip{A}}^k$.

\subsection{Properties of $\mathsf{FI}_\mathsf{A}$}

We establish some properties of the full-information protocol maps $\mathsf{FI}_\mathsf{A}$. 
For example, in the case of the immediate snapshot (i.e., $\mathsf{A} = \mathsf{IS}$), we have that $\Xi_{\mathsf{FI}_\mathsf{IS}} = \textit{Ch}$.
In Figure~\ref{fig:subdivisions}, we depict the full-information protocol maps corresponding to one iteration of the iterated atomic snapshot ($\mathsf{FI}_\mathsf{AS}$) and iterated collect ($\mathsf{FI}_\mathsf{C}$) in $3$-process systems. 
In contrast with the chromatic subdivision, the geometric realization of the resulting protocol maps do not preserve planarity in $\Delta_2$ and are not subdivision operators.
Nevertheless, they share some key properties with the chromatic subdivision operator.
Specifically, Lemma~\ref{lemma:FIP_prop_mesh} and Lemma~\ref{lemma:FIP_prop_cap} demonstrate that all communication primitives $\mathsf{A} \in \smset$ induce a mesh-shrinking protocol map $\Xi_{\fip{A}}$~\cite{distCompTopo}. 
Intuitively, this means that in the $k$-th round of the protocol, the shortest path between any pair of vertices strictly increases from one round to the next.
The lemmata below are used in the proofs of Lemma~\ref{lemma:lateResultLowerBound} and Theorem~\ref{theorem:encodings}.

\begin{lemma} \label{lemma:FIP_prop_mesh}
    Let $\I$ be an input complex, then for $\mathsf{A} \in \smset$:
    \[
    \forall v,w \in V(\I), \{v,w\} \notin \Xi_{\fip{A}}(\I)
    \]
    \end{lemma}
    \begin{proof}
    By contradiction, suppose there exists a pair of vertices $v,w \in V(\I)$ such that the edge $ \{v,w\} \in \Xi_{\fip{A}}(\I)$.
    That is, there is a global configuration in $\Xi_{\fip{A}}(\I)$ where processes $\pi(v)$ and $\pi(w)$ read nothing but their own input.
    In particular, both processes did not see each other's inputs.
    By definition, $\fip{A} \in \textit{ITER}$. Since it follows the pseudocode of Algorithm~\ref{alg:encodedFIP}, we have that all processes finish writing their input before reading the shared memory. 
    Thus, if $\pi(v)$ did not read $\pi(w)$'s input, the read operation of $\pi(v)$ started before $\pi(w)$ finished writing its input. 
    And the write operation of $\pi(v)$ finished before the write operation of $\pi(w)$.
    Conversely, as $\pi(w)$ also did not read $\pi(v)$'s input, the write operation of $\pi(w)$ finished before the write operation of $\pi(v)$, which is a contradiction.  
    \end{proof}
    
    \begin{lemma} \label{lemma:FIP_prop_cap}
    %
    For all $\mathsf{A} \in \smset$, $\Xi_{\fip{A}}$ is a strict carrier map.
    \end{lemma}
    
    \begin{proof}

    Let $\sigma \in \Xi_{\fip{A}}(\A) \cap \Xi_{\fip{A}}(\B)$. By definition, $\sigma$ is a final view configuration of $|\sigma|$ processes that can be reached from both $\A$ and $\B$. 
    Since $\sigma \in \Xi_{\fip{A}}(\A)$, the inputs read by processes in $\pi(\sigma)$ must have originated from some initial configuration in $\A$.
    Similarly, as $\sigma \in \Xi_{\fip{A}}(\B)$, the same inputs must also be in $\B$. It follows that the initial input configuration of $\sigma$ is contained in $\A \cap \B$, implying that $\sigma \in \Xi_{\fip{A}}(\A \cap \B)$. 
    Thus, $\Xi_{\fip{A}}(\A) \cap \Xi_{\fip{A}}(\B) \subseteq \Xi_{\fip{A}}(\A \cap \B)$.

    Conversely, let $\sigma \in \Xi_{\fip{A}}(\A \cap \B)$. This means $\sigma$ is a final configuration that results from an initial configuration in $\A \cap \B$.
    Since $\A \cap \B \subseteq \A $ and $ \A \cap \B \subseteq \B$, it follows that $\sigma \in  \Xi_{\fip{A}P}(\A)$ and $\sigma \in  \Xi_{\fip{A}}(\B)$. 
    Thus, $\sigma \in  \Xi_{\fip{A}}(\A) \cap \Xi_{\fip{A}}(\B)$.
\end{proof}

Finally, the following lemma shows that to obtain the degree of a vertex of the input complex after $r$ iterations of $\fip{A}$ with $\mathsf{A} \in \smset$, is equal to the degree of the vertex after $r$ chromatic subdivisions.

\begin{lemma} \label{lemma:lateResultLowerBound}
Let $\I$ be an input complex and let $v \in V(\mathcal{I})$ be a vertex. Then, for all $r \geq 0$ and $\mathsf{A} \in \smset$:
\[
{\mathrm{deg}(\Xi_{\fip{A}}^r(\I),v)} =
{\mathrm{deg}(\mathrm{Ch}^r\ \I,v)}
\]
\end{lemma}
\begin{proof}
We will prove the statement by induction over $r$. 
Fix a vertex $v \in \I$.
Let $p$ be a process such that $\pi(v)=p$.
\textit{Base case.}
As each face of the protocol complex corresponds to a view of the communication primitive, by Lemma~\ref{lemma:viewh}, $\chdiv \I = \Xi_{\fip{IS}}(\I) \subset \Xi_{\fip{A}}$.
Therefore, ${\deg(\Xi_{\fip{A}}(\I),v)} \geq \deg(\chdiv{\I},v)$.
Moreover, by definition of the communication primitives, the configurations in $\Xi_{\fip{A}}(\I) \setminus \chdiv{\I}$ do not contain the vertex $v$, which implies that ${\deg(\Xi_{\fip{A}}(\I),v)} \leq \deg(\chdiv{\I},v)$.
Thus, ${\deg(\Xi_{\fip{A}}(\I),v)} = \deg(\chdiv{\I},v)$.
Note that by Lemma~\ref{lemma:FIP_prop_mesh}, any edge adjacent to $v$ in $\I$ does not appear in $\Xi_{\fip{A}}(\I)$ nor in $\chdiv{\I}$.

For the inductive step, suppose that ${\mathrm{deg}(\Xi_{\fip{A}}^r(\I),v)} =
{\mathrm{deg}(\chdiv^{\!r}{\I},v)}$. 
It suffices to observe that the new edges generated in the $(r+1)$ subdivision correspond to reachable executions of $\fip{A}$, since, by Lemma~\ref{lemma:FIP_prop_mesh}, neither $\Xi_{\fip{A}}^{r+1}(\I)$ nor $\mathrm{Ch}^{r+1}(\I)$ contains edges which are also in $\Xi_{\fip{A}}^{r}(\I)$ or $\mathrm{Ch}^{r}(\I)$.
As in the base step, the new edges correspond to the central vertices yielded at the subdivision of each face containing vertex $v$.
By a similar reasoning to the base step it follows that the new edges are exactly those in the chromatic subdivision. 
\end{proof}

Therefore, any algorithm A $\in \mathit{ITER}$ that its protocol map is equivalent, to a full-information protocol map must also satisfy the properties stated in Lemma~\ref{lemma:FIP_prop_mesh}, Lemma~\ref{lemma:FIP_prop_cap} and Lemma~\ref{lemma:lateResultLowerBound}.

\section{Necessary Conditions for Full-Information Protocols}
\label{sec:necessary}
We now determine conditions an algorithm $A \in \iter$ must satisfy in order to induce a protocol complex isomorphic to the full-information protocol complex of $\Xi_{\fip{A}}$.
The key concept in achieving this equivalence is \textit{process distinguishability}. 
That is, a process is able to determine the state of another by reading the value published by that process in the communication primitive. 
Importantly, it is not necessary for processes to explicitly share all of their knowledge; a small, decodable value may suffice to uniquely identify a state within the protocol complex.
Exploiting this, we can define \textit{bounded} encoding functions.
The following definitions define the notion of distinguishability for subcomplexes.

\begin{definition}[Vertex distinguishability] \label{def:disting}
    Let $\K$ be a simplicial complex and $\omega : V(\I)\rightarrow E$. We say that $v \in V(\K)$ is \textit{distinguishable} in $\K$ \textit{under} $\omega$ if no adjacent vertex to $v$ has an adjacent vertex that uses the same encoding as $v$ and has the same color as $v$: 
    $ \forall u \in V(\K), \ \{v,u\} \in \K\implies \nexists x \in V(\K)\setminus\{v\} : \{u, x\} \in \K \land \omega(v) = \omega(x) \land \pi(x)=\pi(v)$. 
\end{definition}

\begin{definition}[Simplicial subcomplex distinguishability] \label{def:disting_simplex}
Let $\K$ be a simplicial complex, $\B \subseteq \K$ subcomplex, and $\omega : V(\K)\rightarrow E$. We say that $\B$ is distinguishable in $\K$ under $\omega$ if every vertex $v \in V(\B)$ is distinguishable in $\K$ under $\omega$.
\end{definition}

In other words, Definition~\ref{def:disting} stipulates that a vertex in the protocol complex is distinguishable if all adjacent vertices to the former can tell apart the vertex by just reading the shared value in the communication primitive.
Then, if this relation is true for all vertices on a subcomplex, we say that the whole subcomplex is distinguishable.

Theorem~\ref{theorem:encodings} gives a necessary condition for an algorithm $A \in \iter$ to be equivalent to its full-information counterpart $\fip{A}$. 
At a high level, each face of the input complex must be made distinguishable in $\I$ at some round of $A$.
To make a face $\sigma \in \I$ distinguishable, the encoding function has to assign distinct values to each vertex in $\sigma$ so that, if the input configuration is $\sigma$, all processes can unambiguously map the observed values to the corresponding states in $\sigma$.
The sequence of encoding functions of $A$, $\omega_A$, must account for each global input configuration and convey the same information as in the full-information protocol.
We treat each face $\sigma \in \I$ as a subcomplex in $\I$ when reasoning about distinguishability.

\begin{theorem} \label{theorem:encodings}
Let $\I$ be an input complex, let $A$ be an algorithm in $\iter$ using a communication primitive $\mathsf{A} \in \smset$, and $\{\omega_k\}_{k\leq r}$ its sequence of encoding functions.
Then, for all $r \geq 0$:
\[
    \Xi_{A(r)}(\I) \cong \Xi_{\fip{A}}(\I) \implies \forall \sigma \in \I\; \exists k \leq r\ :\ \sigma\ \text{is distinguishable in}\ \I\ \text{under}\ \omega_{k}
\]
\end{theorem}
\begin{proof}
By contradiction, suppose that there exists a face $\sigma \in \I$ that is not distinguishable in $\I$ under any encoding function in $\{\omega_k\}_{k\leq r}$.
As $\Xi_{A(r)}(\I) \cong \Xi_{\fip{A}}(\I)$, by Lemma~\ref{lemma:FIP_prop_cap} we have that for any subcomplex $\A\subseteq\I$, $\Xi_{A(k)}(\sigma \cap \A) = \Xi_{A(k)}(\sigma) \cap \Xi_{A(k)}(\A)$. 
We want to show that this requires that $\sigma$ is distinguishable in $\I$ under some $\omega_k \in \{\omega_k\}_{k\leq r}$.

Consider the case where $\I$ is a simplicial complex consisting of two simplices $\alpha$ and $\beta$ of dimension $n$ such that $\alpha \cap \beta = \sigma$ and $dim(\sigma)=n-1$. Note that both $\alpha$ and $\beta$ have a vertex that is not in $\sigma$. Let $a$ and $b$ be such vertices respectively and let $p$ be the process assigned to both, i.e., $p=\pi(a)=\pi(b)$.
Let $p$ write the same value for both states, that is: $\omega_k(a) = \omega_k(b)$. Since $a$ and $b$ are adjacent to any vertex in $\sigma$, it follows that $\alpha$ is not distinguishable in $\I$ under $\omega_k$. Analogously, $\beta$ is neither distinguishable in $\I$ under $\omega_k$.
By contradiction, suppose $\Xi_{A(k)}(\sigma) = \Xi_{A(k)}(\alpha) \cap \Xi_{A(k)}(\beta)$. Consider the execution where the global input configuration is $\alpha$ and a process $p'\neq p$ reads the value written by $p$ and $p$ reads nothing but its own input. Let $c_{p'}$ be the resulting final state of process $p'$. Because $\omega_k(a) = \omega_k(b)$, from the perspective of $p'$, the execution is \emph{indistinguishable} from the execution where process $p$ is in state $b$. 
Thus, the vertex $c_{p'}$ is both in $\Xi_{A(k)}(\alpha)$ and $\Xi_{A(k)}(\beta)$.
However, $a \notin \sigma$ and $b \notin \sigma$. Thus, $c_{p'} \notin \Xi_A(\sigma)$, a contradiction.

Therefore, there exists a $\omega_k \in \{\omega_k\}_{k\leq r}$ such that $\sigma$ is distinguishable in $\I$ under $\omega_k$, which contradicts the initial hypothesis.
\end{proof}

Theorem~\ref{theorem:encodings} demonstrates that in an iterative protocol, each facet is made distinguishable at some round $k$. This is a defining property of the iterative protocol maps $\Xi_A$: they must subdivide the facets of the input complex at some step. Indeed, making a facet distinguishable is necessary for the protocol to make progress.
In the case that we want an algorithm $A \in \iter$ to simulate one iteration of full-information in a single round, Theorem~\ref{theorem:encodings} directly implies that the encoding function of $A$ has to make the whole input complex distinguishable.

%

\subsection{Distinguishability graph of a Simplicial Complex}

\begin{figure}
    \centering
    \includegraphics[width=0.6\columnwidth]{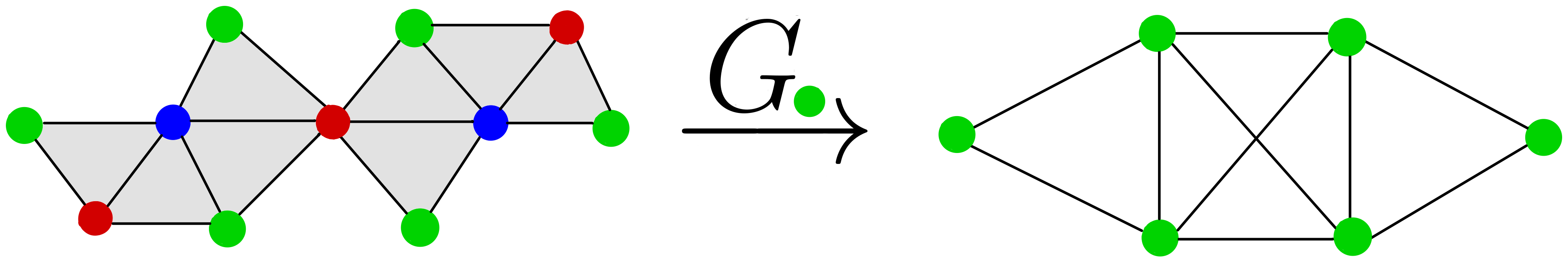}
    \caption{Example of a simplicial complex along with its corresponding distinguishability graph of the green-labeled process. The relative positions of the green vertices are preserved in the drawing.} 
    \label{fig:disting_graph}
\end{figure}

We will further develop the requirements for an encoding function $\omega_k$ to ensure distinguishability of a subcomplex.
To accomplish this, we define the \textit{distinguishability graph}. 
The idea is that, given a simplicial complex, we construct a graph that describes the vertices that must have a different encoding in order for each vertex for the subcomplex to be distinguishable under $\omega_k$. 
Intuitively, the distinguishability graph consists of the set of vertices of the simplicial complex of some particular color, and there is an edge between two vertices if there exists a third vertex in the simplicial complex which is adjacent to both.
Figure~\ref{fig:disting_graph} illustrates a simplicial complex and its associated distinguishability graph for one of the processes.

\begin{definition}[Distinguishability graph]
Let $\mathcal{I}$ be a simplicial complex and $p \in \Pi$ be a process.
We define the \textbf{distinguishability graph} $G_p : \mathcal{I} \times \Pi \rightarrow (V(\mathcal{I}),V(\mathcal{I})\times V(\mathcal{I}))$  of $\mathcal{I}$ with respect to process $p$ as follows:
for each pair of distinct vertices $v,w \in V(\mathcal{A})$ such that $\pi(v)=\pi(w)=p$,
\[
(v,w) \in G_p(\mathcal{A}) \iff \exists t\in V(\mathcal{A}) : v,w \in V(\mathrm{St}(\mathcal{A},t))
\]
\end{definition}

The following theorem establishes an equivalence between determining if an encoding function renders a subcomplex distinguishable and the problem of finding a proper vertex coloring for the distinguishability graphs associated with a simplicial complex.

Formally, given a graph $G=(V,E)$ and $S \subseteq V$. We say that a coloring $\omega$ is proper on $S$ iff $\forall  v \in S,\; \nexists t \in V : (v,t)\in E \land \omega(v)=\omega(t)$.

\begin{theorem} \label{theorem:vertex colouring}
Let $\mathcal{I}$ be a simplicial complex, $\B \subseteq \K$, and $\omega$ an encoding function.
Then,
\[
\mathcal{B} \text{ is distinguishable in } \I  \text{ under } \omega \iff \forall p \in \Pi,\; \omega \text{ is a proper vertex coloring on } V(\B) \text{ in } G_p(\mathrm{St}(\I,\mathrm{St}(\I,\B))). 
\]
\end{theorem}
\begin{proof}
We will prove the necessary and sufficient conditions by contradiction. For the former, suppose that there is an edge $(v,w) \in G_p(\mathrm{St}(\mathrm{St}(\I,\B),\I))$ with $v \in V(\B)$ such that $\omega(v)=\omega(w)$ and $\pi(v)=\pi(w)=p$.
Because such an edge exists, by definition of distinguishability graph, it implies that there exists $t \in V(\mathrm{St}(\mathrm{St}(\I,\B),\I))$ such that $t$ is adjacent to $v$ and $w$ in the protocol complex.
However, as both vertices have the same value under $\omega$, $t$ cannot distinguish $v$ and $w$. A contradiction, as we assumed that $\mathcal{B}$ is distinguishable under $\omega$. 
Now for the sufficient condition, suppose that $\mathcal{B}$ is not distinguishable in $\I$ but $\omega$ is a proper vertex coloring on $V(\B)$ of $G_p(\mathrm{St}(\I,\B),\I))$. 
That implies that there is a vertex $t$ that is adjacent to $w$ and $v$ such that $\omega(v)=\omega(w)$. As $w,v \in \mathrm{St}(\mathcal{I},t)$ we have that $(w,v)$ in $G_p(\mathrm{St}(\mathrm{St}(\I,\B),\I))$. A contradiction, because both vertices have the same color in $\omega$.
\end{proof}

As an outcome of Theorem~\ref{theorem:vertex colouring}, we can now provide insights into the number of different values the encoding function requires by leveraging well-known results from graph theory.
For instance, the number of distinct encodings required to satisfy Theorem~\ref{theorem:encodings} corresponds to the maximum chromatic number (i.e., the smallest number of colors needed to color the vertices of a graph so that no two adjacent vertices share the same color) among all graphs in $\{G_p(\mathcal{I})\}_{p\in\Pi}$. 
The encoding function does not have to explicitly encode any information about the process state but a distinct value that renders the process state distinguishable within the protocol complex.
Then, the transition function, $\sigma_k$, can decode the encoded representation obtained in the $k$-layer view.

\begin{corollary}\label{coro:ima_lowbound}
    Let $\mathcal{I}$ be a simplicial complex, let $\B \subseteq \I$, $\omega : V(\mathcal{I})\rightarrow E$ an encoding function and let $\kappa(G)$ be the size of the largest clique in a graph $G$.
    Then,
    \[
    \mathcal{B} \text{ is distinguishable in } \mathcal{I} \text{ under } \omega \implies |E| \geq \max_{p \in \Pi}{\kappa(G_p(\mathcal{B}))}
    \]
\end{corollary}
\begin{proof}
If $\mathcal{B}$ is distinguishable in $\I$ under $\omega$, then by Theorem~\ref{theorem:vertex colouring}, $\omega$ is a proper vertex coloring in $V(\B)$ of the family of graphs $G_\Pi(\mathrm{St}(\mathrm{St}(\I,\B),\I))=\{G_p(\mathrm{St}(\mathrm{St}(\I,\B),\I))\}_{p\in\Pi}$, which implies it is a proper vertex coloring of all graphs in $G_\Pi(\B)$.
Let $\kappa$ be the size of the biggest clique of the graphs in $G_\Pi(\mathcal{B})$. 
In order to color a clique of size $\kappa$, we need at least $\kappa$ different colors. 
As a result, the cardinality of the output domain --- $|E|$ --- has to be at least as big as $\kappa$.
\end{proof}

\begin{corollary}\label{coro:ima_highbound}
    Let $\mathcal{I}$ be a simplicial complex and $\B \subseteq \I$ a subcomplex. 
    Let $\Delta(G)$ be the biggest vertex degree in a graph $G$.
    Then, there exists an encoding function $\omega : V(\I)\rightarrow E$ such that $\B$ is distinguishable in $\I$ under $\omega$ and $|E| \leq \max_{p \in \Pi}{\Delta(G_p(\mathrm{St}(\mathrm{St}(\I,\B),\I)))}+1$
\end{corollary}
\begin{proof}
    Let $\omega$ be an arbitrary encoding function, in order for $\mathcal{B}$ to be distinguishable in $\I$ under $\omega$, we require $\omega$ to be a proper vertex coloring on $V(\B)$ in all graphs $\{G_p(\mathrm{St}(\mathrm{St}(\I,\B),\I))\}_{p\in\Pi}$.
    By Brook's theorem~\cite{brooks_1941}, the chromatic number of a graph $G$ is less or equal to $\Delta(G)+1$. That means that there exists a proper vertex coloring function $\omega$ whose size of its encoding set $|E|\leq \Delta(G) + 1$. 
    Let $\delta_p$ be a proper coloring on $V(\B)$ in $G_p(\mathrm{St}(\mathrm{St}(\I,\B),\I)))$, because each $G_p$ takes the vertices of different processes, we can define the encoding function $\omega$ by taking each $\delta_p$ for the vertices of a process $p$.
    Then, we have that $\omega$ is a proper coloring on $V(\B)$ in all the graphs $\{G_p(\mathrm{St}(\mathrm{St}(\I,\B),\I))\}_{p\in\Pi}$ and by Theorem~\ref{theorem:vertex colouring}, $\B$ is distinguishable in $\I$ under $\omega$. 
    Moreover, by Brook's theorem $|E| \leq \Delta(G_p(\mathrm{St}(\mathrm{St}(\I,\B),\I))) + 1$.
\end{proof}

\subsection{Distinguishable Subcomplex}

Distinguishability graphs characterize the constraints that an encoding function must satisfy in order to make an arbitrary subcomplex distinguishable.
By Theorem~\ref{theorem:encodings}, an algorithm $A \in \iter$ may span multiple rounds and consequently have multiple encoding functions. 
Therefore, different subcomplexes of the input complex may become distinguishable at different rounds of the execution.
To capture this incremental progress, we introduce the notion of a distinguishable subcomplex.
Intuitively, this subcomplex contains the faces of the input complex whose vertices have already become distinguishable under the sequence of encoding functions used by the protocol. 

\begin{definition} \label{def:disting_subcomplex}
Let $\I$ be an input complex, and let $\omega_A=\{\omega_{k}\}_{1 \leq k \leq r}$ denote the sequence of encoding functions of an algorithm $A \in \iter$.
We define the \emph{distinguishable subcomplex of $\I$ under $\omega_A$}, denoted $\D(\omega_A,\I)$, as follows:
$$
\D(\omega_A,\I) = \bigcup_{1 \leq k < r}{\{ \sigma \in \text{Faces}(\I) : \sigma\ \text{is distinguishable in}\ \I\ \text{under}\ \omega_{k}\}}
$$
\end{definition}

Observe that $\D(\omega_A,\I)$ is a well-defined simplicial subcomplex, as if a face $\sigma \in \D(\omega_A,\I)$ is distinguishable under some $\omega_{A(k)}$, all its faces $\sigma' \subseteq \sigma $ are also distinguishable under the same encoding function. 
Using this definition we can restate Theorem~\ref{theorem:encodings} as follows: 

\begin{corollary} \label{corollary:distset}
Let $\I$ be an input complex, let $A$ be an algorithm in $\iter$ using a communication primitive $\mathsf{A} \in \smset$. Then,
\[
\Xi_{A}(\I) \cong \Xi_{\fip{A}}(\I) \Rightarrow \D(\omega_A,\I) = \I
\]
\end{corollary}

With the above definitions, our objective is to study algorithms \(A \in \iter\) together with their associated sequence of encoding functions $\omega_A$ where each encoding function $\omega_k \in \omega_A$ satisfies $|E_k| \le 2^{b}-1$.
Furthermore, by Corollary~\ref{corollary:distset}, any such sequence must eventually render the entire input complex distinguishable, i.e., \(\D(\omega_A,\I) = \I\).

\subsection{Computational Hardness of Finding Optimal Encodings}

It is natural to seek, for a given input complex under a bounded encoding constraint, a sequence of encoding functions that is simultaneously optimal in the number of rounds (i.e., achieving round-optimal full-information simulation) and in the size of the encoding set (i.e., minimizing the number of bits communicated per round).

However, it follows directly from Theorem~\ref{theorem:vertex colouring} that computing an encoding function with minimum output domain is $\NP$-Hard, as it is equivalent to determining the chromatic number of the induced distinguishability graph. 
Note that the hardness arises from the input complex, which can be arbitrarily large.

\begin{proposition}
Let $\mathcal{I}$ be simplicial complex and $\B \subseteq \I$. 
The optimization problem of finding a function $\omega : V(\I) \rightarrow E$ such that $\mathcal{B}$ is distinguishable in $\I$ under $\omega$, with no other function $\omega' : V(\I) \rightarrow E'$ satisfying $|E'| < |E|$ is $\NP$-hard.
\end{proposition}
\begin{proof}
    It suffices to consider the case where $\B = \I$.
    We will show that given a graph $H$, we can build a chromatic input complex $\mathcal{H}$, such that $G_{p_0}(\mathcal{H})=H$. Thus, the set of distinguishability graphs is the set of all graphs.
    For each edge $(v_i,v_j) \in H$, we build the simplicial complex $\mathcal{H}_{ij}=\{\{(p_0,v_i),(p_1,v_{ij})\},$
    $\{(p_1,v_{ij}),(p_0,v_j)\},\{(p_0,v_i)\},\{(p_1,v_{ij})\},\{(p_0,v_j)\},\emptyset\}$. Then we build $\mathcal{H}$ by doing the union of all these simplices: $\mathcal{H}=\bigcup_{(v_i,v_j) \in H}\mathcal{H}_{ij}$. That is, we build a 2-process simplicial complex by adding a pair of edges for each edge in $H$. Then, we have that $(p_0,v_i),(p_0,v_j) \in V(\mathrm{Lk}(\mathcal{H},(p_1,v_{ij})))\ \forall i,j$. As these are the only vertices in the link, we have that $H=G_{p_0}(\mathcal{H})$.
    
    As a result, the optimization problem requires determining the chromatic number of arbitrary big graphs, which is $\NP$-Hard~\cite{Karp-NP}.
\end{proof}

Determining optimal encoding functions is already computationally intractable for a single round. 
A similar difficulty arises when considering sequences of encodings across multiple rounds.
In particular, the problem of finding a shortest sequence of encoding functions that satisfies the above constraints is also $\NP$-Hard.

\begin{proposition}\label{NP}
Let $\I$ be an input complex, and $b>0$.
Let $A$ be an algorithm in $\iter$, $\omega_A$ its corresponding sequence of encoding functions and $E_k$ the output domain of each encoding function $\omega_k \in \omega_A$.
The problem of finding an algorithm $A \in \iter$ satisfying $\forall \omega_k \in \omega_A,\; |E_k|\leq 2^b -1$ and $\D(\omega_A,\I) = \I$, such that there is no other algorithm $A' \in \iter$ with a shorter sequence $\omega_{A'}$ satisfying the former conditions is $\NP$-hard.
\end{proposition}
\begin{proof}
We will prove this by showing that the problem can be used to solve an arbitrary instance of the Set Cover problem, thereby showing it is NP-hard via a reduction from Set Cover~\cite{Karp-NP}.

Consider encoding functions whose image is contained in $\{1,\bot\}$.
Let $A(\mathcal{I})$ be an algorithm that solves the optimization problem stated in the proposition, returning a valid cover with the respective encodings used.    
Let $U$ and $S\subseteq 2^U$ be a set and a collection of subsets, respectively.
We build the complex $\mathcal{I}$ as follows: for each $u \in U$, define a facet $\Delta_{|U|}$ such that for each $s\in \overline{S}$ (complement of set $S$), the facets with labels in $s$ share a common vertex. As we have $|U|$ facets, and each simplex could be attached to $|U|$ different simplices, $\mathcal{I}$ is well-defined.

Now $A(\mathcal{I})$ will return a minimal cover such that the facets of each subset will not share a vertex (as they could not be made distinguishable by encodings writing only 1). Thus, the labels of the facets of each subset are in $S$, and each facet corresponds to an element $u \in U$. It follows that $A(\mathcal{I})$ solves the optimization version of Set Cover.
\end{proof}

We have shown that computing an optimal sequence of encoding functions is \(\NP\)-hard. 
We therefore focus on the asymptotic complexity of encoding schemes.
In particular, we derive lower bounds on round and bit complexity as functions of $r$, $b$, and $n$.

\subsection{Combinatorial Characterization of the Necessary Condition}

We now link the necessary condition of Lemma~\ref{lemma:lateResultLowerBound} to a combinatorial property of the input complex and subsequently characterize its asymptotic growth as a function of the number of full-information rounds, processes, and available bits.

\begin{lemma} \label{lemma:min_rounds}
Let $\I$ be an input complex, and let $A \in \mathit{ITER}$ be an algorithm using a communication primitive $\mathsf{A} \in \smset$  such that $\Xi_{A}(\I) \cong \Xi_{\fip{A}}(\I)$ and a sequence of encoding functions $\omega_A$.
Then,
\[
|\omega_A| \geq 
\Big\lceil \frac{\max_{p \in \Pi}
{\kappa(G_p(\I))}}
{2^b - 1}
 \Big\rceil \geq
\Big\lceil \frac{\max_{v \in V(\I)}
{\mathrm{deg}(\I,v)}}
{n (2^b - 1)}
 \Big\rceil
\]
\end{lemma}
\begin{proof} 
Since $\Xi_A(\I) \cong \Xi_{\fip{A}}(\I)$, Corollary~\ref{corollary:distset} implies that the sequence of encodings $\omega_A$ must satisfy $\D(\omega_A,\I)=\I$. In particular, the encoding functions must collectively render every vertex of $\I$ distinguishable.
Fix a process $p \in \Pi$. By Corollary~\ref{coro:ima_lowbound}, making $\I$ distinguishable requires that, within any clique of $G_p(\I)$, all vertices are assigned distinct encoding values.
Hence the total number of distinct encodings that must be assigned to vertices of process $p$ is at least $\kappa(G_p(\I))$.
Each encoding function $\omega_k$ is constrained to $b$ bits, and therefore its image contains at most $2^b-1$ values. Consequently, a single encoding function can assign distinct encodings to at most $2^b-1$ vertices of a given process.
It follows that a sequence of encoding functions must have length at least
$
|\omega_A| \geq
\Big\lceil
\frac{\max_{p\in\Pi}\kappa(G_p(\I))}
{2^b-1}
\Big\rceil
$.
Moreover, observe that any clique in $G_p(\I)$ arises from a vertex $t\in V(\I)$ whose star contains several vertices of color $p$. Indeed, if $(v,w)$ is an edge of $G_p(\I)$, by definition there exists $t$ such that both $v$ and $w$ belong to $\mathrm{St}(\I,t)$. 
Thus a clique corresponds to a set of vertices of the same process that all appear in the star of a common vertex.
Let $v\in V(\I)$ be a vertex of maximum degree $D=\deg(\I,v)$. The star $\mathrm{St}(\I,v)$ contains $D$ adjacent vertices distributed among the $n=|\Pi|$ processes. 
In the worst case these vertices are evenly distributed among processes, so that at least $D/n$ of them have the same process label. These vertices form a clique in the corresponding distinguishability graph.
\end{proof}

\section{Analysis of the $f$-vector of Iterative Subdivisions}
\label{sec:fvector}
Corollaries~\ref{coro:ima_lowbound} and~\ref{coro:ima_highbound} relate the number of distinct values required by an encoding function to combinatorial properties of the indistinguishability graphs, namely their \emph{clique number}, i.e., the number of vertices in a maximum clique, and its \emph{degree}, i.e., the maximal number of edges incident to its vertices.
Then, Lemma~\ref{lemma:min_rounds} gives a lower bound on the necessary number of encoding functions by the maximal vertex degree of the input complex. 
In order to derive asymptotic bounds on the communication complexity of such algorithms, it therefore suffices to understand how these quantities evolve as the protocol complex grows under iterated executions of the full-information protocol.

By Lemma~\ref{lemma:lateResultLowerBound}, the degree of a vertex in the protocol complex obtained after $r$ iterations of $\fip{A}$ coincides with its degree in the $r$-th iteration of the chromatic subdivision of the input complex.
Consequently, the necessary conditions of protocols in $\iter$ can be studied through purely combinatorial properties of iterated applications of the chromatic subdivisions to the input complex. 

\paragraph{Additional standard constructions on simplicial complexes.}
Given a complex $\K$, the \emph{open star} of a subcomplex $\mathcal{S}\subseteq\K$, $\mathrm{St}^\circ(\K,\mathcal{S})$ is defined as: $\mathrm{St}^\circ(\K,\mathcal{S})=\{\sigma \in \K : \mathcal{S} \subseteq \sigma \}$. 
Note that the open star does not yield a complex but a subset of $\K$.
An $(n−1)$-face in $\K$ is a boundary face if it is a face of exactly one facet of $\K$.
The boundary subcomplex of $\K$, denoted $\partial(\K)$, is the
set of faces contained in the $(n-1)$ boundary faces in $\K$.
%
Then we can define the interior of $\K$ as $Int\  \K=\K \setminus \partial(\K)$.
By definition, the boundary of $\K$ is a complex but the interior is not.
%
%
%
We define the interior of an open star as $Int\ \ostar{\K}{\S} :=\; (Int\,\K)\ \cap\ \mathrm{St}^\circ(\K,\mathcal S)$~\footnote{Throughout this section, we work with abstract simplicial complexes, and all the additional constructions are defined combinatorially (i.e., in terms of face inclusion rather than a geometric realization). 
In particular, our definition of the open star is purely combinatorial, differing from the topological point-set one.
}.
The simplicial join of two complexes $\K$ and $\mathcal{B}$, denoted as $\K * \mathcal{B}$ is the complex with set of vertices $V(\K)\cup V(\mathcal{B})$ whose faces are all the unions $\alpha \cup \beta,\alpha \in \K$ and $\beta \in \mathcal{B}$.
Given a non-negative integer $l$, the $l$-skeleton of a complex $\K$, denoted as $Skel^l(\K)$, is the set of simplices of $\K$ with dimension at most $l$. In particular, $Skel^0(\K)=V(\K)$ and $Skel^n(\K) = \K$.

\paragraph{$f$-vector.}
The \textit{$f$-vector} of an $n$-dimensional complex $\K$ is $f(\K)=(f_{-1}(\K),f_0(\K),\ldots,f_n(\K))$, where $f_k(\K)=|\{\sigma \in \K : dim(\sigma)=k\}|$ for $k \in [0,n]$. 
We adopt the convention that $f_{-1}(\K)=1$ if $\K \neq \emptyset$ and $f_{-1}(\K)=0$ otherwise.
That is, $f_k(\K)$ enumerates the $k$-dimensional faces of $\K$. 
We extend this definition to arbitrary subsets $S \subseteq \K$ (not necessarily subcomplexes) by setting $f_k(S)=|\{\sigma \in S : \dim(\sigma)=k\}|$ and using the same convention for $f_{-1}(S)$.
In particular, this defines the $f$-vector of an open star.
For instance, $f_{k}(\Delta_n) = \binom{n+1}{k+1}$ and $f_k(\ostar{\Delta_n}{v}) = \binom{n}{k}$, for $v \in V(\Delta_n)$ and $k \in [0,n]$.

In particular, the degree of a vertex $v$ in a simplicial complex $\I$ is determined by the number of vertices in its star, $\deg(\I,v) = |V(\mathrm{St}(\I,v))|-1$,
which corresponds to the $f_1$ entry of the $f$-vector of the star of $v$. 
Thus, bounding the communication complexity reduces to analyzing the growth of the $f$-vector of stars of vertices under iterated chromatic subdivision.

\begin{theorem}[$f$-vector of a subdivision] \label{th:fk_tauA}
Let $\tau$ be a subdivision operator, and let $\K$ be an $n$-dimensional complex. Then, for every $k \in [0,n]$,
\[
f_k(\tau(\K)) = \sum_{i=k}^{n}{{f}_i(\mathcal{K})\ {f}_k(Int\ \tau(\Delta_i))}
\]
\end{theorem}
\begin{proof}
    Fix $k \in [0,n]$ and let $\K$ be some $n$-dimensional complex.
    We prove the identity by induction over the skeleton of $\mathcal{K}$. 
    For $i \in [0,n]$, let $S(i)={f}_k(\tau(Skel^i \K))$.
    Recall that $Skel^n \K = \K$.
    We want to show $S(i)=S(i-1)+f_i(\K)\cdot f_k(Int\ \tau(\Delta_i))$. 
    %
    %
    \textit{Base case}.
    By definition, $dim(Skel^i \K) \leq i$, and as $\tau$ is a subdivision operator, $dim(\tau(Skel^i \K)) \leq i$.
    Then, if $i\leq k$, $f_k(\tau(Skel^i \K))=0$.
    \textit{Inductive step}.
    Define $K:= \tau(Skel^i \K) - \tau(Skel^{i-1} \K)$.
    Observe that $K$ is a set containing subsets of $V(\K)$ of size $i+1$.
    Then, we have that $K \cup \tau(Skel^{i-1} \K) = \tau(Skel^i \K)$ and $K \cap \tau(Skel^{i-1} \K) = \emptyset$. 
    Thus, $f_k(K) + f_k(\tau(Skel^{i-1} \K)) = f_k(\tau(Skel^i \K))$.

    From the inductive hypothesis we have that $f_k(\tau(Skel^{i-1} \K)) = S(i-1)$. 
    %
    %
    Because $\tau$ is a subdivision operator, each element of $K$ is generated from subdividing an $i$-face of $\K$ with $\tau$.
    As there are $f_i(\K)$ $i$-faces in $\K$, we have to enumerate the $k$-faces yielded from subdividing each $i$-face of $\K$ with $\tau$.
    Let $\sigma \in Skel^i \K$ be an $i$-face, note that $\sigma$ itself is a simplex of dimension $i$.
    It remains to observe that $f_k(\tau(\sigma)) = f_k(\tau ( \partial (\sigma))) + f_k(\tau ( Int\ \sigma))$.
    Where $f_k(\tau ( \partial (\sigma)))$ is already comprised in $S(i-1)$, as $\tau ( \partial (\sigma)) \subseteq \tau(Skel^{i-1} \K)$. 
    Therefore, any new face not enumerated in $S(i-1)$ is in $ Int\ \tau (\sigma)$. 
    However, by definition of interior and $\tau$, for any $\sigma' \in Skel^i \K$ such that $\sigma' \neq \sigma$, we get $Int\ \tau(\sigma) \cap \tau(\sigma') = \emptyset$. 
    Thus for the purpose of enumerating faces, we can replace $\sigma$ with $\Delta_i$.
    As we have $f_i(\K)$ $i$-faces, the new $i$-faces enumerated in $S(i)$ yielded in the subdivision are $f_i(\K)\cdot f_k(Int \ \tau (\Delta_i)) $.
    By solving the now proven recurrence relation, we get the final summation expression.
\end{proof}

Theorem~\ref{th:fk_tauA} shows that the $f$-vector of $\tau(\K)$ is completely determined by the $f$-vector of $\K$ together with the values $f_k(Int,\tau(\Delta_i))$, i.e., by how $\tau$ subdivides the standard simplices.
Lemma~\ref{lemma:f_k_tau_star} derives a similar formula for an open star, obtained by applying the theorem only to those faces of $\K$ that contain a fixed vertex $v \in V(\K)$.

\begin{lemma} \label{lemma:f_k_tau_star}
Let $\tau$ be a subdivision operator, let $\K$ be an $n$-dimensional complex, and fix $v \in V(\K)$.
For each $i \in [0,n]$, fix a vertex $v_i \in V(\Delta_i)$.
Then, ${f}_0(\mathrm{St}^\circ(\tau(\mathcal{K}),v)) = 1$ and for every $k \in [1,n]$,
\[
{f}_k(\mathrm{St}^\circ(\tau(\mathcal{K}),v)) = \sum_{i=k}^{n}{{f}_i(\mathrm{St}^\circ(\mathcal{K},v))\ {f}_k(Int\ \mathrm{St}^\circ(\tau(\Delta_i),v_i))}
\]
\end{lemma}
\begin{proof}
    First note that $f_0(\mathrm{St}^\circ(\tau(\K),v))=1$, since the only $0$-face of $\tau(\K)$ that contains $v$ is the vertex $v$ itself.
    Now fix $k\geq 1$.
    The proof is identical to that of Theorem~\ref{th:fk_tauA} after restricting every set to the open star at $v$.
    Concretely, in the skeleton induction we replace $\tau(Skel^i\K)$ by $\mathrm{St}^\circ(\tau(Skel^i\K),v)$ and the difference $\tau(Skel^i\K) - \tau(Skel^{i-1}\K)$ by $\mathrm{St}^\circ(\tau(Skel^i\K),v)-  \mathrm{St}^\circ(\tau(Skel^{i-1}\K),v)$.
    Then the same argument shows that new $k$-faces arise only from subdividing $i$-faces $\sigma$ with $v\in\sigma$, and as $\tau$ is a subdivision operator, each new $k$-face has a unique carrier.
    Hence, each $i$-face contributes $f_k\!\bigl(Int\  \mathrm{St}^\circ(\tau(\Delta_i),v_i)\bigr)$ $k$-faces, and there are $f_i(\mathrm{St}^\circ(\K,v))$ such $i$-faces.
\end{proof}

\begin{lemma} \label{lemma:fIntStCh}
Let $v \in V(\Delta_n)$, and fix $v_i \in V(\Delta_i)$.
Then, for every $k\geq 0$,
\[
{f}_k(Int\ \mathrm{St}^\circ(\mathrm{Ch}\ \Delta_n,v)) = \sum_{i=1}^{k}{\binom{n}{i} {f}_{k-i}(\mathrm{St}^\circ(\mathrm{Ch}\ \Delta_{n-i},v_i))}
\]
\end{lemma}
\begin{proof}
We aim to enumerate the $k$-faces in $Int\ \mathrm{Ch}\ \Delta_n$ which contain the vertex $v$.
For $k=0$, by definition of interior, the right-hand side is an empty sum, hence equals $0$, which agrees with the left-hand side since ${f}_k(Int\ \mathrm{St}^\circ(\mathrm{Ch}\ \Delta_n,v)) = 0$.
Since these faces are in $Int\ \mathrm{Ch}\ \Delta_n$, they must necessarily include at least one vertex of different color than $v$ from the corresponding facet in the central simplex.
We note this facet as $\sigma$.
Hence, the number of faces will be all possible simplicial joins of $v$ with this facet and then with faces on the boundary of $\mathrm{St}^\circ(\mathrm{Ch}\ \Delta_n,v)$.

Let $k > 1$ (by definition of interior, if $k = 0$, we have ${f}_0(Int\ \mathrm{St}^\circ(\mathrm{Ch}\ \Delta_n,v))=0$).
Suppose we want to construct a $k$-face from an $(i-1)$-face of $\sigma$. 
To do this, we take a face $\delta \in \sigma$ such that $dim(\delta)=i-1$ and perform a simplicial join with $v$: $\delta * v$.
Note that we can choose $\binom{n}{i}$ different $(i-1)$-faces. 
We perform the simplicial join $\delta * v$ and obtain an $i$-face in the interior. Now, if we want a $k$-face, we have no other option but to perform a second simplicial join with another face of dimension $(k-i)$ on $\partial\  \mathrm{St}^\circ(\mathrm{Ch}\ \Delta_n,v)$.
The faces we can join with $\delta$ are those that do not contain any of the colors already in $\delta$; these are found in the subdivision of one of the $(k-i)$-faces of $\Delta_n$.
Note that, as it is a subdivision, if we restrict ourselves to one of these faces, it is equal to the subdivision of a simplex of the dimension of such face. Therefore, the faces we can join are ${f}_{k-i}(\Delta_{n-i})$. To obtain all faces, we have to enumerate from taking a single vertex from $\sigma$ ($dim(\delta * v)=1$) to taking a $k-1$ face ($dim(\delta * v)=k$), thus obtaining the formulated equation.
\end{proof}

Lemma~\ref{lemma:f_k_tau_star} holds for an arbitrary subdivision operator and any simplicial complex.
Instantiating $\tau$ to the chromatic subdivision and applying it iteratively yields the corresponding identities for the iterated chromatic subdivision.
Corollary~\ref{corollary:fk_chir_a} and Corollary~\ref{corollary:fk_st_chi_delta} directly follow from Lemma~\ref{lemma:f_k_tau_star} and Lemma~\ref{lemma:fIntStCh}.

\begin{corollary} \label{corollary:fk_chir_a}
Let $\K$ be an $n$-dimensional complex, for each $i \in [0,n]$, fix $v_i \in V(\Delta_i)$ and fix $v \in V(\K)$.
Then, for every $r \geq 1$, ${f}_0(\mathrm{St}^\circ(\mathrm{Ch}^r\ \mathcal{K},v)) =1$, and for every $k \in [1,n]$,
\[
{f}_k(\mathrm{St}^\circ(\mathrm{Ch}^r\ \mathcal{K},v)) = \sum_{i=k}^{n}{{f}_i(\mathrm{St}^\circ(\mathrm{Ch}^{r-1}\ \mathcal{K},v))\sum_{j=1}^{k}{\binom{i}{j} {f}_{k-j}(\mathrm{St}^\circ(\mathrm{Ch}\ \Delta_{i-j},v_{i-j}))}}
\]
\end{corollary}

\begin{corollary}\label{corollary:fk_st_chi_delta}
Fix $v \in V(\Delta_n)$, for each $i \in [0,n]$, fix $v_i \in V(\Delta_i)$. Then, ${f}_0(\mathrm{St}^\circ(\mathrm{Ch}\ \Delta_n,v))=1$ and for every $k \in [1,n]$,
\[
 {f}_k(\mathrm{St}^\circ(\mathrm{Ch}\ \Delta_n,v)) = \sum_{i=k}^n{\binom{n}{i} \sum_{j=1}^k{\binom{i}{j} {f}_{k-j}(\mathrm{St}^\circ(\mathrm{Ch}\ \Delta_{i-j},v))}}
\]
\end{corollary}

\begin{lemma} \label{lemma:fIntCh}
For every $k\geq 0$,
\[
f_k(Int\ \chdiv{\Delta_n}) = \sum_{i=0}^k{\binom{n+1}{i+1}f_{k-i-1}(\chdiv{\Delta_{n-i-1}})}
\]
\end{lemma}
\begin{proof}
We follow the same reasnoning as in the proof of Lemma~\ref{lemma:fIntStCh} but we consider all faces contained in $ Int\ \mathrm{Ch}\ \Delta_n$.
Let $\alpha$ be the central $n$-simplex of $\mathrm{Ch}\ \Delta_n$.
By definition, $\alpha \subseteq  Int\ \mathrm{Ch}\ \Delta_n$.
Observe that a face $\sigma$ is in $Int\ \mathrm{Ch}\ \Delta_n$ if and only if $\sigma \cap \alpha \neq \emptyset$.
Otherwise, $\sigma$ is yielded from the subdivision of some facet of $\Delta_n$ and is contained in the boundary $\partial \chdiv{\Delta_n}$.
Fix $k\geq 0$.
We want to enumerate the $k$-faces in $Int\ \mathrm{Ch}\ \Delta_n$.
Any such faces contains at least one vertex in $\alpha$. 
Thus, a $k$-faces $\sigma$ in  $Int\ \mathrm{Ch}\ \Delta_n$ can be decomposed as $\sigma = \delta \ast \beta$, where $\delta \subseteq \alpha$, $dim(\delta)=i$, $dim(\beta)=k-i-1$ and $i \leq k$.
Note that if $dim(\beta)=-1$, $\beta = \emptyset$.
There are $\binom{n+1}{i+1}$ choices for such an $i$-face $\delta\subseteq \alpha$.
Fix $\delta$.  
Then, the set of labels of vertices of $\beta$, has to be disjoint from $\delta$ and $\beta \subseteq \partial \chdiv{\Delta_n}$.
In particular there is a unique face $F \subseteq \partial {\Delta_n}$ which a disjoint vertex labeling from $\delta$.
Therefore, $\delta$ can be joined with any of the faces spanned by $\chdiv{F} \cong \chdiv{\Delta_{n-i-1}}$.
Thus, there are $f_{k-i-1}(\Delta_{n-i-1})$ faces which can be joined to obtain a face in the interior. 
Considering taking $\delta$ with $\dim(\delta)=i$ with $i \in [0,k]$ yields the claimed identity.
\end{proof}

\begin{corollary} \label{corollary:fk_chir_a2}
Let $\K$ be an $n$-dimensional complex.
Then, for every $r \geq 1$ and for every $k \in [0,n]$,
\[
{f}_k(\mathrm{Ch}^r\ \mathcal{K}) = \sum_{i=k}^{n}{{f}_i(\mathrm{Ch}^{r-1}\ \mathcal{K})\sum_{j=0}^{k}{\binom{i+1}{j+1} {f}_{k-j-1}(\mathrm{Ch}\ \Delta_{i-j-1})}}
\]
\end{corollary}

\begin{corollary}\label{corollary:fk_chi_delta2}
Fix $v \in V(\Delta_n)$, for each $i \in [0,n]$, fix $v_i \in V(\Delta_i)$. Then, ${f}_0(\mathrm{St}^\circ(\mathrm{Ch}\ \Delta_n,v))=1$ and for every $k \in [1,n]$,
\[
{f}_k(\mathrm{Ch}\ \Delta_n) = \sum_{i=k}^{n}{\binom{n+1}{i+1}\sum_{j=0}^{k}{\binom{i+1}{j+1} {f}_{k-j-1}(\mathrm{Ch}\ \Delta_{i-j-1})}}
\]
\end{corollary}

Lemma~\ref{lemma:argmax_chst} shows that, under iterative subdivision, the maximal value of $f_k(\mathrm{St}^\circ(\tau^r(\K),v))$ over all vertices $v \in V(\tau^r(\K))$ is attained at a vertex already present after the first subdivision.

\begin{lemma} \label{lemma:argmax_chst}
Let $\tau$ be a subdivision operator and $\mathcal{K}$ be an $n$-dimensional complex.
For every $r \geq 1$ and every $k \leq n$, every vertex
$v \in V(\tau^r(\K))$ that maximizes
$f_k(\mathrm{St}^\circ(\tau^r(\K),v))$
belongs to $V(\tau(\K))$.
\end{lemma}
\begin{proof}
We will proceed by induction on $r$ for an arbitrary $k$. The base step $r=1$ is immediate. For the inductive step, let $v=\argmax_{v \in V(\tau^r(\K))}{{f}_k(\mathrm{St}^\circ(\tau^r(\K),v))}$. By inductive hypothesis, $v \in V(\tau(\K))$. We have that $\forall \Tilde{v} \in V(\tau^r(\K)), {f}_k(\mathrm{St}^\circ(\tau^r(\K),\Tilde{v}))\leq{f}_k(\mathrm{St}^\circ(\tau^r(\K),v))$. From Lemma~\ref{lemma:f_k_tau_star}, we have that ${f}_k(\mathrm{St}^\circ(\tau^{r+1}(\K),v))$ is computed as a sum of the faces in $\mathrm{St}^\circ(\tau^r(\K),v)$ multiplied by a constant that come from the simplex subdivision ${f}_k(Int\ \mathrm{St}^\circ(\tau(\Delta_i),v))$. Thus, $\forall \Tilde{v} \in V(\tau^r(\K)), {f}_k(\mathrm{St}^\circ(\tau^{r+1}(\K),\Tilde{v}))$ $\leq {f}_k(\mathrm{St}^\circ(\tau^{r+1}(\K),v))$. Next, the star of the new vertices $V(\tau^{r+1}(\K))\setminus V(\tau^r(\K))$ will have a constant ${f}$-vector, irregardless of $r$, based on the dimension of the face they are a central simplex of. Hence, the vertex with the biggest ${f}$-vector (in all its coordinates) of its open star will still be $v$.
\end{proof}

\subsection{Asymptotic Analysis on the $f$-vector of the Iterated Chromatic Subdivision}

Using the $f$-vector equations derived in this section, we can perform an asymptotic analysis w.r.t the number of processes in the system (directly corresponding to the dimension of the input complex) and the number of applications $r$ of the chromatic subdivision. 
Thus, our objective will be to find tight asymptotic bounds for ${f}_k(\mathrm{St}^\circ(\mathrm{Ch}^r\ \mathcal{A},v))$ and ${f}_k(\mathrm{Ch}^r\ \mathcal{A})$ for $k \leq n$.
%

The following combinatorial identity allows us to collapse the double summation in the recurrence for $f_k(\mathrm{St}^\circ(\mathrm{Ch}\ \Delta_n,v))$ into a closed form, which is key to the asymptotic analysis in Lemma~\ref{lemma:ThetaBoundFkdelta}.

\begin{lemma} \label{lemma:A1}
Let $n,k,r \in \mathbb{N}$ such that $r\leq r \leq n$, and let $(n)_k$ denote the falling factorial.
Then,
\[
\sum_{i=k}^{n}{\binom{n}{i}\binom{i}{r} b^{i-\alpha} (i-r)_{k-r} }
=
\frac{b^{k-\alpha}}{r!}(b+1)^{n-k}(n)_k
\]
\end{lemma}
\begin{proof}
The result follows from the binomial theorem and the factorial formulation of the binomial coefficient:
  \begin{multline*}
    \sum_{i=k}^{n}{\binom{n}{i}\binom{i}{r} b^{i-\alpha} (i-r)_{k-r} } = \sum_{i=k}^{n}{\frac{n!}{(n-i)! r! (i-k)!} b^{i-\alpha} } 
    = \frac{n!}{r!} \frac{(n-k)!}{(n-k)!} \sum_{i=0}^{n-k}{\frac{b^{i+k-\alpha}}{(n-k-i)! i!} } \\
    = \frac{(n)_k}{r!} b^{k-\alpha} \sum_{i=0}^{n-k}{\binom{n-k}{i} b^i}
    = \frac{b^{k-\alpha}}{r!}(b+1)^{n-k}(n)_k 
  \end{multline*}
\end{proof}

\begin{lemma} \label{lemma:ThetaBoundFkdelta}
Let $v \in V(\Delta_n)$ be a vertex in $\Delta_n$.
Then, for any $k \leq n$,
\[
{f}_k(\mathrm{St}^{\circ}(\mathrm{Ch}\ \Delta_n,v) \in \Theta \Big(\frac{(k+1)^{n-k} (n)_k }{\ln (2)^{k-1}}\Big)
\]
\end{lemma}
\begin{proof}
First, to ease notation, let $T(k,n) := {f}_k(\mathrm{St}^{\circ}(\mathrm{Ch}\ \Delta_n,v)) $, and define the bounding function $f(k,n) := \frac{(k+1)^{n-k} (n)_k }{\ln(2)^{k+1}}$. It is important to observe that in order to compute $T(k,n)$, we require the values $T(k',n') : k'\in [0,k-1]$ and $n' \in [0,n-1]$. Consequently, we can prove the result through a general induction over $n$ for all $k$.

The base case is immediate: $T(0,n) \overset{\mathrm{def}}{=} 1 \in \Theta(1) = \Theta(f(0,n)), \forall n \in \mathbb{N}$. For the induction step, we will prove it by computing the limit of $\frac{T(k,n)}{f(k,n)}$ as $n$ goes to infinity for all $k\leq n$ and showing this limit converges to a non-zero constant.  

\begin{multline*}
    {f}_k(\mathrm{St}^{\circ}(\mathrm{Ch}\ \Delta_n,v))  \in \Theta(f(k,n)) \overset{\text{by IH}}{\iff} \sum_{i=k}^{n}{\binom{n}{i} \sum_{j=1}^k {\binom{i}{j}\frac{ (k-j+1)^{i-k} (i-j)_{k-j}}{\ln (2)^{k-j-1}}}} \\
    \overset{\text{Lemma~\ref{lemma:A1}}}{=} \sum_{j=1}^{k}{\frac{\ln (2)^j}{\ln (2)^{k-1} j!} (k-j+2)^{n-k} (n)_k}
    =\frac{(n)_k}{\ln (2)^{k-1}}\sum_{j=1}^{k}{\frac{(k-j+2)^{n-k} \ln (2)^j}{j!}}  \in  \Theta(f(k,n)) \\
    \overset{\forall k\leq n}{\iff} \lim_{n \to \infty}{\frac{\frac{(n)_k}{\ln (2)^{k-1}}\sum_{j=1}^{k}{\frac{(k-j+2)^{n-k} \ln (2)^j}{j!}}}{(n)_k (k+1)^{n-k} \ln (2)^{-k+1}}} = C > 0
    \overset{\forall k\leq n}{\iff} \lim_{n \to \infty}{\sum_{j=1}^k{(\frac{k-j+2}{k+1})^{n-k} \frac{\ln (2)^j}{j!}}}=C>0 
\end{multline*}

We will now show that the limit of the last equation is in the interval $(\ln 2, 1)$ for all values of $k\leq n$. For the lower bound, noting that the term inside the sum is strictly positive, we take the first term of the sum: 
\[
\lim_{n \to \infty}{\sum_{j=1}^k{\big(\frac{k-j+2}{k+1}\big)^{n-k} \frac{\ln (2)^j}{j!}}} > \lim_{n \to \infty} (\frac{k+1}{k+1})^{n-k} \frac{\ln 2}{1!} = \ln 2 > 0 \ \ \forall{k\leq n}
\]
For the upper bound we use the fact that $\sum^{\infty}_{j=1}{\frac{\ln (2)^j}{j!}}=1$ and $\frac{k-j+2}{k+1} \leq 1 \ \forall k \leq n$. Thus the following series is a multiplication of a convergent sequence by another one which their values are in $(0,1]$:
\[
\lim_{n \to \infty}{\sum_{j=1}^k{\big(\frac{k-j+2}{k+1}\big)^{n-k} \frac{\ln (2)^j}{j!}}} < \lim_{n \to \infty}{\sum_{j=1}^{\infty}{\big(\frac{k-j+2}{k+1}\big)^{n-k} \frac{\ln (2)^j}{j!}}} < 1 \ \ \forall{k\leq n}
\]

Consequently, the asymptotic ratio between $\frac{T(k,n)}{f(k,n)}$ is in the interval $(\ln 2, 1)$ $\forall k\leq n$. It follows that $f(k,n)$ is a tight asymptotic bound of $T(k,n)$

\end{proof}

Now that we have a tight asymptotic bound on ${f}_k(\mathrm{St}^{\circ}(\mathrm{Ch}\  (\Delta_n),v))$, we can now give a bound for the equation in Lemma~\ref{lemma:fIntStCh}. Which will be useful for computing the final result regarding the ${f}$-vector of the iterative chromatic subdivision:

\begin{lemma} \label{lemma:bound_intch}
For every $k \in [1,n]$,
\[
\sum_{j=1}^k{\binom{n}{j}{f}_{k-j}(\mathrm{St}^{\circ}(\mathrm{Ch}\  \Delta_{n-j},v))} \in \Theta \Big(\frac{(k+1)^{n-k} (n)_k }{\ln (2)^{k-1}}\Big)
\]
\end{lemma}
\begin{proof}
Defining $T(k,n)$ as in Lemma~\ref{lemma:ThetaBoundFkdelta}, let $R(k,n):= \sum_{j=1}^k{\binom{n}{j}T(k-j,n-j)}$ and $f(k,n) := \frac{(k+1)^{n-k} (n)_k }{ln(2)^{k+1}}$.
\begin{multline*}
    R(k,n) \in \Theta(f(k,n)) \overset{\text{Lemma~\ref{lemma:ThetaBoundFkdelta}}}{\iff} \sum_{j=1}^{k}{\binom{n}{j} \frac{(k-j+1)^{i-k} (i-j)_{k-j}}{\ln (2)^{k-j-1}}} \\
    =\frac{(n)_k}{\ln (2)^{k-1}}\sum_{j=1}^{k}{\frac{(k-j+1)^{n-k} \ln (2)^j}{j!}}  \in  \Theta(f(k,n)
    \overset{\forall k\leq n}{\iff} \lim_{n \to \infty}{\sum_{j=1}^k{(\frac{k-j+1}{k+1})^{n-k} \frac{\ln (2)^j}{j!}}}=C>0 \ 
\end{multline*}
Where the previous limit is the same as in Lemma~\ref{lemma:ThetaBoundFkdelta} but with $k-j+1$ instead of $k-j+2$ on the numerator and $\frac{k-j+1}{k+1} < 1$. Thus the limit is non-zero convergent.
\end{proof}

We are now ready to give a tight asymptotic bound on the iterative chromatic subdivision of a simplicial complex. 

\begin{theorem}[Asymptotic bound on the iterative chromatic subdivision] \label{theorem:finalBound}
Let $\mathcal{A}$ be an $n$-dimensional simplicial complex, $v \in V(\mathcal{A})$ and $r>1$. Then,
\[
{f}_k(\mathrm{St}^\circ(\mathrm{Ch}^r\ \mathcal{A},v)) \in \Theta \bigg(\Big(\frac{n!}{\ln (2)^{n-1}}\Big)^{r-1}\frac{(k+1)^{n-k}(n)_k}{\ln(2)^{k+1}} \bigg)
\] 
\end{theorem}
\begin{proof}
    Again, to relieve notation we write $P(k,r) := {f}_k(\mathrm{St}^\circ(\mathrm{Ch}^r\ \mathcal{A},v))$, $T$ and $R$ defined as in the proof of Lemma~\ref{lemma:bound_intch}. Then we denote the bounding function $f(k,r) := \bigg(\frac{n!}{\ln (2)^{n-1}}\bigg)^{r-1}\frac{(k+1)^{n-k}(n)_k}{\ln(2)^{k+1}}$. Note that $n$ is implicitly defined as the dimension of $\mathcal{A}$. Now instead of doing induction over $n$, we will prove it by induction over $r$ for all $k \leq n$.

    The base step: 
    \begin{align*}
        P(k,2) \in \Theta(f(k,2)) &\overset{\text{Lemma~\ref{lemma:bound_intch}}}{\iff} \sum_{i=k}^n{P(i,1) \frac{(k+1)^{i-k} (i)_k}{\ln (2)^{k-1}}} \in \Theta(f(k,2)) \\
        &\iff \sum_{i=k}^n{\frac{(i+1)^{n-i}(n)_i(k+1)^{i-k}(i)_k}{\ln (2)^{k-1} \ln(2)^{i-1}}} \in  \Theta(f(k,2))
    \end{align*}

    Note that for all $k$, the asymptotic dominating term of the sum w.r.t $n$ is the last term of the summation when $i=n$. Thus $P(k,2) \in \Theta(f(k,2))$. Now for the inductive step:

    \begin{multline*}
    P(k,r) \in \Theta(f(k,r)) \overset{\text{by IH}}{\iff} \sum_{i=k}^n{f(k,r-1) \sum_{j=1}^k{\binom{i}{j} T(k-j,i-k) }} \in \Theta(f(k,r))  \\
    \overset{\text{Lemma~\ref{lemma:bound_intch}}}{=}\sum_{i=k}^n{f(i,r-1)\frac{(k+1)^{i-k} (i)_k}{\ln (2)^{k-1}}} 
    = \sum_{i=k}^n{\bigg(\frac{n!}{ln(2)^{i-1}}\bigg)^{r-2}\frac{(k+1)^{i-k} (i)_k}{\ln(2)^{k-1}}}  \in \Theta(f(k,r)) \\
    \end{multline*}

    Observe that the dominating term of the sum is again given by $i=n$ wich concludes the proof:

    \[
    \bigg(\frac{n!}{\ln(2)^{n-1}}\bigg)^{r-2}\frac{n!}{ln(2)^{n-1}}\frac{(k+1)^{n-k}(n)_k}{\ln(2)^{k+1}} = f(k,r)
    \]

\end{proof}

An interesting fact is that ${f}_n(\mathrm{St}^{\circ}(\mathrm{Ch}\ \Delta_n,v))$ is an approximation of the Fubini numbers~\cite{fubiniNumbers}: ${f}_n(\mathrm{St}^{\circ}(\mathrm{Ch}\  \Delta_n,v)) \sim \frac{n!}{2 \ln (2)^{n+1}}$, which corresponds to the tight asymptotic bound of Theorem~\ref{theorem:finalBound}.

\begin{theorem} \label{theorem:finalasympv2}
Let $\mathcal{A}$ be an $n$-dimensional simplicial complex. For every $r \ge 1 $and every $k \le n$,
\[
f_k(\mathrm{Ch}^r\mathcal{A})
\in
\Theta \Bigg(
\left(\frac{n!}{\ln(2)^{\,n-1}}\right)^{r-1}
\frac{(k+1)^{\,n-k}(n)_k}{\ln(2)^{\,k+1}}
\Bigg)
\]
\end{theorem}
\begin{proof}

From Corollary~\ref{corollary:fk_chir_a2},
$
f_k(\mathrm{Ch}^r\mathcal{A})
=
\sum_{i=k}^{n}
f_i(\mathrm{Ch}^{r-1}\mathcal{A})
\sum_{j=0}^{k}
\binom{i+1}{j+1}
f_{k-j-1}(\mathrm{Ch}\Delta_{i-j-1}).
$.
Define $G(k,i)=
\sum_{j=0}^{k}
\binom{i+1}{j+1}
f_{k-j-1}(\mathrm{Ch}\Delta_{i-j-1})$.
Using Lemma~\ref{lemma:fIntCh} together with the asymptotic bound for $f_k(\mathrm{Ch}\Delta_m)$, we obtain
$
G(k,i)
\in
\Theta\!\left(
\frac{(k+1)^{\,i-k}(i)_k}{\ln(2)^{\,k}}
\right)
$.
Substituting into the recurrence gives
$
f_k(\mathrm{Ch}^r\mathcal{A})
\in
\Theta\!\left(
\sum_{i=k}^{n}
f_i(\mathrm{Ch}^{r-1}\mathcal{A})
\frac{(k+1)^{\,i-k}(i)_k}{\ln(2)^{\,k}}
\right)
$
For fixed k, the dominant contribution comes from the largest index $i=n$. Hence
$
f_k(\mathrm{Ch}^r\mathcal{A})
\in
\Theta\!\left(
f_n(\mathrm{Ch}^{r-1}\mathcal{A})
\frac{(k+1)^{\,n-k}(n)_k}{\ln(2)^{\,k}}
\right).
$
Finally, the number of $n$-simplices created by one chromatic subdivision satisfies
$
f_n(\mathrm{Ch}\Delta_n)
\sim
\frac{n!}{\ln(2)^{\,n-1}},
$
so iterating the subdivision yields
$
f_n(\mathrm{Ch}^{r}\mathcal{A})
\in
\Theta\!\left(
\left(\frac{n!}{\ln(2)^{\,n-1}}\right)^r
\right)
$.
\end{proof}

\section{Asymptotic Complexity of Bounded Full-Information Protocols}
\label{sec:results}

By applying the asymptotic bounds from Theorem~\ref{theorem:finalBound} to the construction in Lemma~\ref{lemma:lateResultLowerBound}, we derive a lower bound on the round complexity as a function of the number of available bits $b$ per round for an algorithm $A \in \iter$.

\begin{theorem}
\label{theorem:final_low}
Let $\I$ be an $n$-dimensional input complex and let $r>0$. 
Let $A \in \mathit{ITER}$ be an algorithm that uses a communication primitive $\mathsf{A} \in \smset$, let $\omega_A$ be its corresponding sequence of encoding functions and $E_k$ the output domain of each encoding function $\omega_k \in \omega_A$ satisfying $|E_k| \leq 2^b-1$.
Then, if $\Xi_{A}(\I) \cong \Xi_{\fip{A}}^r(\I)$, algorithm $A$ must perform $\Omega\Bigl({(n!)^{r-1} \cdot 2^{n-b}} \Bigr)$ rounds.
\end{theorem}

In the context of Theorem~\ref{theorem:final_low}, if algorithm $A \in \iter$ performs a single round per iteration of the full-information protocol, then $\Omega( r n \log n)$ is a lower bound on the bit complexity of $A$ for $n>1$.
Conversely, if $b=2$, we have that $\Omega((n!)^{r-1} 2^n)$ is a lower bound on the round complexity of $A$. 
For a two-process system, Theorem~\ref{theorem:final_low} yields a constant $\Theta(1)$ round complexity for any $b$.

In the case of simulating an iteration of full-information in a single round, one can define an encoding function in which a unique value is assigned to each vertex of the input complex.
In such case, the asymptotic bit complexity corresponds to the number of vertices at $r$ applications of the standard chromatic subdivision on the input complex.
Then by Corollary~\ref{theorem:final_low} and Theorem~\ref{th:trivial_upper} we obtain a tight bound of $\Theta(r n \log n)$ on the bit complexity.

\begin{theorem} \label{th:trivial_upper}
    Let $\I$ be an input complex.
    Let $n>2$ be the number of processes in the system and $r \geq 1$.
    There exists an algorithm $A \in \iter$ using a communication primitive $\mathsf{A} \in \smset$ such that $\Xi_A(\I) \cong \Xi_{\fip{A}}^r(\I)$, $|\omega_A|=1$, with bit complexity $O(r n \log n)$.
\end{theorem}
\begin{proof}
    Fix $r \geq 1$.
    It suffices to define an encoding function $\omega_k : V(\Xi_{\fip{A}}^k(\I)) \rightarrow E_k$ for the resulting protocol complex at each iteration $k$ of the full-information protocol for $0 \leq k < r$.
    We define $\omega_k$ as to assign a different value to each vertex of the protocol complex.
    By definition, for each $k$, $\Xi_{\fip{A}}^k(\I)$ is distinguishable under $\omega_k$ for every $k$..

    Let $v_p, w_q \in V(\Xi_{\fip{A}}^k(\I))$. Because $v_p$ distinguishes $w_q$, there exists a locally invertible function $\omega^{-1}_k$ for each state of $p$ such that $\omega^{-1}_k(\omega_k(w_q),v_p)=w_q$. 
    Thus if the state transition function in Algorithm~\ref{alg:encodedFIP}, $\sigma$ has the form $v,s \mapsto \omega^{-1}_k(v,s)$, then $\sigma$ is injective in $v$ and returns the state of each vertex in $\mathrm{St}(\mathcal{I},v_p)$. 
    Next, if the values written by $\omega_k$ makes all pair of vertices distinguishable, then, from all states in $\mathcal{I}$, the behavior of $\Xi_A$ is the same: from a view of the communication primitive, it decodes the views of the other processes, obtaining their respective local states.
    Thus, $\Xi_A(\mathcal{I})$ has the same behavior as the \emph{full-information} protocol $\Xi_{\fip{A}}$: from reading the communication primitive we get the views of the other processes. Thus $\Xi_A(\mathcal{I}) \cong \Xi_{\fip{A}}(\mathcal{I})$.

    Finally, the bit complexity of $\omega_k$ is precisely $O(\log f_0(\Xi_{\fip{A}}^r(\I)))$, which by Theorem~\ref{theorem:finalasympv2} and the fact that all faces in $\Xi_{\fip{A}}^r(\I) \setminus \chdiv^{\!r}{\I}$ do not contain vertices in $\I$, yields
    $O\big(\log (n!/(\ln{2})^n)^{r-1}\big)= O(r n \log n)$. 
\end{proof}

%
The case of two processes is special: a constant bit complexity is required to simulate any iteration of the full-information protocol. 
We defer the formulation of such bounded protocol to Section~\ref{sec:2p}.
In particular such a protocol yields an algorithm that solves $\epsilon$-agreement for arbitrarily small $\epsilon$ with $2$-bit complexity.

It is noteworthy that the asymptotic complexity depends only on the number of processes in the system and the number of iterations of the full-information protocol. 
The input size (the number of simplices of the input complex) plays only a constant factor when calculating the $f$-vector over iterative subdivisions. 
The number of indistinguishable states only increases \emph{locally} within each simplex.

\subsection{Greedy Star: an Asymptotically Optimal Algorithm} 
\label{sec:greedy_star}

In this section we define a construction for obtaining protocols in $\iter$ which are asymptotically optimal in their round complexity and their encoding functions are bounded by some arbitrary number of $b\geq2$ bits.

In Algorithm~\ref{alg:greedyCover}, we describe an algorithm --- motivated by the inequality of Lemma~\ref{lemma:min_rounds} --- to construct a sequence of encoding functions $\omega_\mathsf{GS}$.
We devise an algorithm to construct a sequence of encoding functions that yield \emph{distinguishable subcomplexes} whose union covers of the input complex.
We call this algorithm Greedy Star, as it proceeds by iteratively selecting vertices in $\I$ and ensuring that each selected vertex is distinguishable by assigning it a unique encoding within its star.
To maintain distinguishability, the algorithm enforces that the star of each selected vertex has an empty intersection with the stars of previously selected vertices within the same encoding round.

As a result, we can construct algorithms $\A \in \mathit{ITER}$ --- with a communication primitive $\mathsf{A} \in \smset$ --- using Greedy Star to obtain their encoding function $\omega_A$. 
By construction, Greedy Star is deterministic and computed locally by each process.
Thus, each process shares the same sequence $\omega_A$ of encoding functions.
Then, the next state function $\sigma_{A}$ will map the read values to the respective processes' states by inverting the encodings of $\omega_{A}$. 
Note that by definition of distinguishability, such an inverse exists, as all functions in $\omega_{A}$ are injective within each star. 
In each round, processes accumulate information about the states of other processes which access the communication primitives, provided that the cover subcomplex for that round includes the vertex representing their current state.
In the final round $r$, each process gains the same knowledge as it would in a single iteration of the full-information protocol (Theorem~\ref{theorem:gs_gives_collect}). 
Consequently, the terminal states in the protocol complexes  $\Xi_A$ and $\Xi_{\fip{A}}$ are effectively identical. 
This protocol complex can then be used as input for a subsequent iteration of the full-information protocol.

\begin{algorithm}
\SetAlgoLined
\KwIn{Chromatic input complex $\mathcal{I}$.}
\KwOut{Sequence of encoding functions $\{\omega_r\}_{r\geq 1}$ such that $\D(\{\omega_r\}_{r\geq 1},\I)=\I$.}

$\mathcal{A} \gets \emptyset$; $V \gets V(\mathcal{I})$; $W \gets \emptyset$; $r \gets 1$ \\
\While{$\mathcal{A}\neq\mathcal{I}$}{
    \textbf{let} $\omega_r : V(\I) \rightarrow \mathbb{N}\cup\{\bot\},\ w_r(x) = \bot$\\
    $\mathcal{U} \gets \emptyset$ \\
    $V_u \gets \emptyset$ \\
    \ForEach{$v \in V$}{
        $\mathcal{S} \gets \mathrm{St}(\mathcal{I},v)$ \\
        \If{$\forall w \in V(\mathcal{S}), \mathrm{St}(\mathcal{I},w) \cap \mathcal{U} = \emptyset$}{
            $\mathcal{U}\gets\mathcal{U} \cup \mathcal{S}$ \\
            $V_u \gets V_u \cup V(\mathcal{S})$ \\ 
            \tcp{redefine $\omega_r$ for vertices in the star.}
            \textbf{let} $\forall x \in V(\mathcal{S}), k \in \mathbb{N},\ \omega_r(x)=k : \nexists y \in V(\mathcal{S}),\ x\neq y \land \omega_r(y)=k \land \pi(y)=\pi(x)$  \\
        }
    }
    $\mathcal{A}\gets\mathcal{A}\cup\ \mathcal{U}$\\
    $V \gets V\setminus V_u$ \\
    $W \gets W \cup \{\omega_r\}$ \\
    $r \gets r+1$ \\
}
\Return $W$
\caption{Greedy Star encoding algorithm: $GS(\mathcal{I})$}
\label{alg:greedyCover}
\end{algorithm}

\begin{lemma}
The Greedy Star Cover algorithm outputs a sequence of encoding functions such that $D(\omega_{\mathsf{GS}}(\I),\I) = \I$.
\end{lemma}
\begin{proof}
First, note that the while condition is not satisfied until $\mathcal{A}=\I$, so we have to show that the condition is eventually reached.
Observe that in each for loop we start with  $\mathcal{U}=\emptyset$, so on each iteration a new star subcomplex will be added to $\mathcal{A}$.
Note that $V$ is initially the set of vertices of $\mathcal{I}$, and vertices are only removed when they are used to generate a new star. 
As $\bigcup_{v \in V(\mathcal{I})}{\mathrm{St}(v)}=\mathcal{I}$, after a finite number of iterations we have that $\A = \I$.
Each star $\mathrm{St}(v)$ added to the cover corresponds to the image of the input complex under one encoding function $\omega_k \in \omega_{\mathsf{GS}}$, and is made distinguishable from the others by construction (since the algorithm ensures that these stars are disjoint in the encoding space by assigning unique encodings).
Hence, for every face $\sigma \in \I$, there exists some $k$ such that $\sigma$ lies in a star $\mathrm{St}(v)$ added at iteration $k$ and is distinguishable in $\I$ under $\omega_k$. Therefore, \(\sigma \in D(\omega_k, \I)\), and so \(\sigma \in D(\omega_{\mathsf{GS}}, \I)\). 
\end{proof}

Algorithms $A \in \textit{ITER}$ can use encoding functions whose output domain cardinality is at most $2^b-1$. 
However, an encoding function $\omega_k \in \omega_{\mathsf{GS}}(\I)$ returned by Greedy Star may have higher cardinality.
Therefore, we decompose each encoding function $\omega_k \in \omega_{\mathsf{GS}}(\I)$ into several encoding functions, each using at most $2^b-1$ non-$\bot$ encodings.
Let $\omega_k : V(\mathcal I) \to E_k$ be an encoding function
produced by Greedy Star. 
We replace $\omega_k$ by a sequence $\omega_k \mapsto (\omega_{k,1},\ldots,\omega_{k,t})$ with $t = \left\lceil \frac{|E_k|}{2^b-1} \right\rceil$.
Let $E_k = E_{k,1} \cup \cdots \cup E_{k,t}$ be any partition such that $|E_{k,i}| \le 2^b-1$ for all $i$. For each $i$, define the encoding function $\omega_{k,i} : V(\mathcal I) \to E_{k,i}$ by
\[
\omega_{k,i}(v) =
\begin{cases}
\omega_k(v) & \text{if } \omega_k(v) \in E_{k,i},\\
\bot & \text{otherwise.}
\end{cases}
\]

Vertices $v$ with $\omega_{k,i}(v)=\bot$ ignore the contents of shared memory during that iteration and preserve their local state.
Replacing each $\omega_k$ with the sequence $(\omega_{k,1},\ldots,\omega_{k,t})$ yields the decomposed sequence $\omega^*_{\mathsf{GS}}(\mathcal I)$.

Operationally, replacing $\omega_k$ by the sequence $(\omega_{k,1},\ldots,\omega_{k,t})$ corresponds to simulating a single round defined by $\omega_k$ using $t$ rounds. 
In round $i$, a process on state $v$ writes $\omega_{k,i}(v)$ to the communication primitive.
Over the t rounds, each value in $E_k$ is written exactly once, so the sequence collectively simulates the behavior of $\omega_k$ under the constraint that each round uses at most $2^b-1$ distinct encodings. \guillermo{ok?}

\begin{lemma}\label{lemma:max_rounds_star}
Let $\mathcal{I}$ be an input complex, $b\geq2$.
Let $\omega^*_{\mathsf{GS}}(\mathcal I)$ be the sequence obtained from $\omega_{\mathsf{GS}}(\mathcal I)$ by decomposing each encoding function $\omega_k$ into encoding functions whose output domain has cardinality at most $2^b-1$.
Then,
\[
|\omega^*_{\mathsf{GS}}| \leq 
 4\Big\lceil \frac{\max_{v \in V(\I)}
{\mathrm{deg}(\I,v)}}
{n (2^b - 1)}
 \Big\rceil
\]
\end{lemma}
\begin{proof}
At each iteration of the outer while loop, the algorithm constructs an intermediate set $\mathcal{U}$ of disjoint vertex stars, each centered at a vertex $v$, and assigns a unique encoding to each facet in these stars.
Since each encoding function can use at most $2^b-1$ non-$\bot$ encodings, a single encoding function can distinguish at most $2^b-1$ facets for a given color.
In a vertex star, facets may be incident to vertices of
different colors, and facets associated with different colors do not require distinct encodings as they correspond to different processes. 
In the best case, the $\deg(\mathcal{I},v)$ facets in the star of $v$ are evenly distributed among the $n$ possible colors.
Thus, at most $\deg(\mathcal{I},v)/n$ facets require distinct
encodings for any given color. 
Consequently, to assign unique encodings to all facets in any selected star it suffices to use
$
M := \Big\lceil \frac{\max_{v \in V(\mathcal{I})} \deg(\mathcal{I}, v)}{n (2^b - 1)} \Big\rceil
$
encoding functions per iteration of the while loop.

Let $\mathcal{A}$ denote the set of facets that have already been assigned encodings (i.e., covered), and let $\overline{\mathcal{A}} = \mathcal{I} \setminus \mathcal{A}$ be the remaining uncovered subcomplex.
Distances are measured in the $1$-skeleton of $\mathcal{I}$, where two vertices are adjacent if they share an edge.

In the first iteration, the stars in $\mathcal{U}$ are disjoint by construction and their union covers a subset of facets of
$\mathcal{I}$. 
Any vertex in $\overline{\mathcal{A}}$ that is adjacent to a vertex in $\mathcal{A}$ is at distance at most $3$ in the $1$-skeleton from a fully encoded facet. 
Otherwise, there would exist a vertex whose star is disjoint from all previously selected stars, contradicting the maximality of $\mathcal{U}$.
In the second iteration, new stars are selected around such frontier vertices. 
Since these vertices are adjacent to already covered regions, their stars contain facets that intersect previously encoded facets and therefore become fully encoded. 
Repeating this reasoning, after the third iteration the remaining uncovered subcomplex $\overline{\mathcal{A}}$ consists only of isolated facets or vertices surrounded by covered stars.
In the fourth iteration, any remaining uncovered facet must intersect the star of a vertex selected in this iteration, and is therefore assigned an encoding. Thus, after at most four iterations of the while loop all facets of $\mathcal{I}$ are covered.
Since each iteration requires at most $M$ encoding functions, the total number of encoding functions in the decomposed sequence satisfies $|\omega^*_{\mathsf{GS}}| \leq 4M$.
\end{proof}

It is important to remark that the Greedy Star algorithm is agnostic to the underlying communication primitive of the algorithm: it only provides a sequence of encoding functions that render the input complex distinguishable. 
In particular, using stronger primitives such as atomic or immediate snapshots with Greedy Star, instead of collect, offers no advantage: the resulting protocol map collapses to that of iterated collect.
This is because each encoding function is decomposed into multiple rounds, isolating communication to disjoint regions of the input complex at each round.
Processes accumulate information across these rounds to simulate a single full-information snapshot iteration.
As a result, the atomicity and immediacy properties of the snapshot primitive are lost, and the overall behavior reduces to that of a sequence of collect operations.
%
Nonetheless, we show that an algorithm using atomic registers as communication primitive, along with the encoding functions produced by Greedy Star, faithfully simulates the unbounded full-information iterated collect protocol.
Thus, By Lemma~\ref{lemma:max_rounds_star} Greedy Star yields an asymptotically round-optimal protocol.

\begin{theorem} \label{theorem:gs_gives_collect}
Let $\mathcal{I}$ be an input complex, and let $GS \in \textit{ITER}$ be an algorithm that uses a communication primitive $\mathsf{A} \in \smset$, with a sequence of encoding functions $\omega_{GS} = GS(\mathcal{I})$ produced by the Greedy Star algorithm. 
Then:
\[
\Xi_{GS}(\mathcal{I}) \cong \Xi_{\fip{C}}(\mathcal{I}),
\]
\end{theorem}
\begin{proof}
    First, we will show that $ \Xi_{GS}(\I) \subseteq \Xi_{\fip{C}}(\I)$. We want to show that any global configuration of ${GS}(\I) $ is also a reachable global configuration in $\fip{C}$.
    By the pseudocode of Algorithm~\ref{alg:encodedFIP}, there is an execution of GS where every process writes and reads only its own input.
    Such execution is present in $\Xi_{\fip{C}}(\I)$.
    %
    %
    Since processes retain all previously read inputs, it suffices to show that any execution of iterated collect can be extended by allowing a process to read additional inputs.
    Consider an execution $\alpha$ of iterated collect and a process $p$ which has not read the input of some other process $q$ in $\alpha$.
    We can construct an execution $\alpha'$ identical to $\alpha$ except that $p$ additionally reads the input of $q$.
    It suffices to rearrange the read operation of $p$ on the register of process $q$ in $\alpha$ such that it occurs afterwards the write operation by $q$.
    Note that the views of all other processes remain unchanged.
    Therefore, any global configuration reachable in $GS$ is also reachable in $\fip{C}$, and we have $\Xi_{GS}(\I) \subseteq \Xi_{\fip{C}}(\I)$. 

    It remains to show that $\Xi_{\fip{C}} (\I) \subseteq \Xi_{GS}(\I)$.
    Consider a final configuration $\sigma \in \Xi_{\fip{C}} (\I)$. 
    Because $GS$ covers the entire input complex, there exists a round $k$ in $GS$ where the input configuration is distinguishable under $\omega_{GS}(k)$.
    Hence, there exists an execution $\beta$ in $GS$ such that the view of all process at round $k$ is exactly the same as in $\sigma$.
    We now construct an extension $\beta'$ of $
    \beta$ in which all processes terminate while preserving the same final configuration $\sigma$.
    Because the model allows arbitrary asynchrony, we can extend the execution by scheduling all remaining steps of processes such that each read after round $k$ returns a value already observed by the reading process. Consequently, no process obtains new inputs beyond what it had at round $k$.
    Thus, no process gains additional information and its local state remains unchanged
    Thus, the final configuration of $\beta'$ is equal to $\sigma$.
    Therefore, $\Xi_{\fip{C}}(\I) \subseteq \Xi_{GS}(\I)$ and $\Xi_{\fip{C}}(\I) \cong \Xi_{GS}(\I)$.
\end{proof}

From a bounded implementation of the iterated collect model, we can derive bounded algorithms with stronger communication patterns, such as atomic and immediate snapshots.
In particular, Algorithm~$5$ from~\cite{delporte2024computational} simulates a single round of immediate snapshot using $n$ rounds of iterated collect.
This simulation can be applied in a black-box manner to our bounded iterated collect implementation, yielding bounded full-information algorithms for snapshot-based models.
As a result, if the bounded iterated collect algorithm terminates in $R$ rounds, the corresponding derived algorithm terminates in $O(n R)$ rounds. 
Therefore, a linear factor in the number of processes is added to the asymptotic round complexity.

Combining Theorem~\ref{theorem:final_low} and Theorem~\ref{theorem:gs_gives_collect}, we derive asymptotic bounds that characterize the fundamental trade-off between the bit complexity ($b$ bits available per process per round) and the round complexity (i.e., the number of rounds) of bounded full-information algorithms.

\begin{theorem}[Tight bound on the bounded full-information iterated collect protocol] \label{theorem:finalbound_IC}
Let $\I$ be an $n$-dimensional input complex, and let $r>0$. 
Let $A \in \mathit{ITER}$ be an algorithm using the collect communication primitive $\mathsf{A}=\mathsf{C}$ using at most $b$ bits per process per round when publishing on the communication primitive. 
Let $\omega_A$ be its associated encoding function.
Then there exists such an algorithm $A$ satisfying $\Xi_{A}(\I) \cong \Xi_{\fip{C}}^r(\I)$, and the number of rounds satisfies: 
$$|\omega_A| \in \Theta\Bigl( (n!)^{r-1} \cdot 2^{n-b}\Bigr)$$
Moreover, no such algorithm exists that uses asymptotically fewer rounds. 
\end{theorem}

\begin{theorem}[Bounds on the bounded full-information iterated snapshot protocols]  \label{theorem:finalbound_Snap}

Let $\I$ be an $n$-dimensional input complex, and let $r>0$.
Let $\fip{A}$ denote the \emph{unbounded full-information protocol} that uses atomic or immediate snapshots as its communication primitive ($A\in\{\mathsf{AS},\mathsf{IS}\}$). 
Let $A \in \mathit{ITER}$ be an algorithm that uses the same communication primitive as $\fip{A}$, but communicates at most $b$ bits per process per round on the communication primitive.
Let $|\omega_A|$ denote the number of encoding functions used by A, i.e., the number of rounds performed.
Then any such algorithm $A$ satisfying $\Xi_{A}(\I) \cong \Xi_{\fip{A}}^r(\I)$ must perform at least
$
|\omega_A| \in \Omega\bigl((n!)^{r-1} \cdot 2^{n - b}\bigr)
$
rounds, and there exists such an algorithm that performs at most
$
|\omega_A| \in O\bigl((n!)^{r-1} \cdot 2^{n - b} \cdot n\bigr)
$ rounds.
\end{theorem}

\section{Case of a Two-process System: Optimal 2-bit Full-Information Protocol}
\label{sec:2p}

All results presented so far apply to systems with three or more
processes ($n \geq 3$). 
In this section we consider the special case of a two-process system.

We show that it is possible to simulate $r$ iterations of the full-information protocol in exactly $r$ rounds using only two bits ($b=2$) per round by presenting an explicit protocol construction with an optimal encoding function and state transition function.
Furthermore, we show how the resulting full-information protocol can be used to solve approximate agreement.

In the two-process approximate agreement task, the processes start with input values $x$ and $y$, respectively, and must output values $x'$ and $y'$ such that $|x' - y'| \le \frac{1}{\epsilon}$, for a given $\epsilon > 0$. 
Moreover, if a process does not receive any information from the other, it must output its own input value. 
The problem was first presented, together with a solution, in~\cite{aproxag} and was shown to be solvable using atomic registers.

For simplicity, we address an inputless version of approximate agreement where one process has input $1$ and the other $0$. In terms of $(\I,\O, \Delta)$, we have that $\mathcal{I}=\{0,1,\{0,1\}\}$.
Without loss of generality, we assume process $p_0$ has input $0$ and $p_1$ input $1$. 
The output graph consists of a path of $\epsilon$ edges, whose vertices are $V(\mathcal{O})=\{(p_0,0),(p_1,\frac{1}{\epsilon}),(p_0,\frac{2}{\epsilon}),\dots,(p_0,\frac{\epsilon-1}{\epsilon}),(p_1,1)\}$. 
The task specification is defined by the carrier map $\Delta((p_0,0))={(p_0,0)}$, $\Delta((p_1,1))={(p_1,1)}$ and $\Delta(\{(p_0,0),(p_1,1)\})=\mathcal{O}$.
Observe that the case where $\epsilon=0$, the task is equivalent to the consensus problem, which is unsolvable in the wait-Free read-write model~\cite{consensus}.

\begin{figure}
    \centering

    \includegraphics[width=0.65\columnwidth]{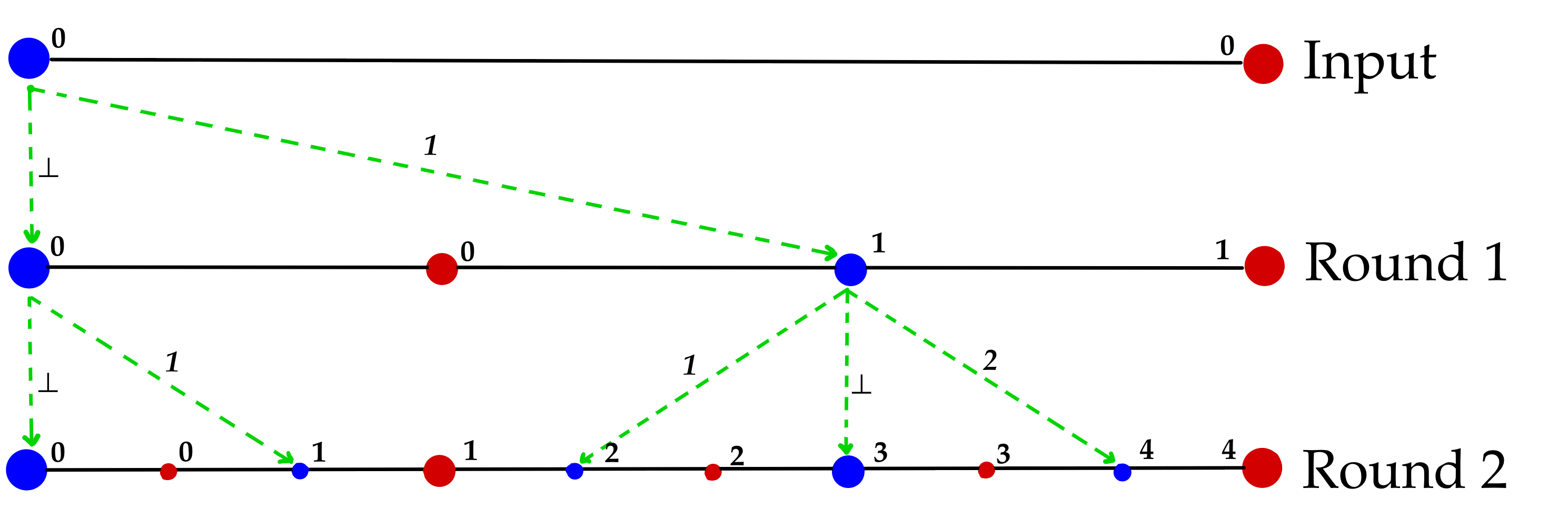}
    \caption{From top to bottom, the input complex, protocol complex for one round, and the protocol complex for two rounds of the bounded approximate agreement algorithm.
    Numbers next to the vertices indicate the internal state of the process.
    Green arrows illustrate the state transition function executed at each round for the blue process.} 
    \label{fig:b2l3}
\end{figure}

Using the full-information protocol, we can achieve approximate agreement with precision $\frac{1}{\epsilon}$ using $\log_3(\ceil{\frac{1}{\epsilon}})$ iterations~\cite{distCompTopo}.
Interestingly, it is sufficient to communicate two bits per process per iteration to solve approximate agreement regardless of $\epsilon$.
The key intuition is illustrated in Figure~\ref{fig:b2l3}.
Given the local view of a process, the state of the other process is restricted to at most two possibilities. Consequently, a process only needs to distinguish between two cases, and thus the communication can be encoded using two distinct non-$\bot$ messages (i.e., two bits).
This property is invariant to the number of rounds.

Next, we present an approximate agreement protocol with 2 bit complexity.
Subsequently, we define the decision function $\delta$, the encoding function $\omega$ and the $\textit{next\_state}$ transition function that encapsulates the internal state of the process at each round.
Recalling the structure of an algorithm in $\iter$ specified in Section~\ref{sec:mymodel}, let the state space $\mathcal{S} \subset \mathbb{N}$ and the next-state function $next\_state : \mathcal{S} \rightarrow \mathcal{S}$ as defined in algorithm~\ref{alg:nextState}.
Then, the encoding function alternates which message to write: $\omega: \mathcal{S} \rightarrow \{1,2\},\  \omega(s)=1+(s \mod 2)$. 
Finally, the decision map $\delta_i : \mathcal{S} \rightarrow \{\frac{k}{\epsilon} : k \in [0,\epsilon]\},\ \delta_i(s)=\frac{2s + i}{\epsilon}$. 
We call the protocol map of this algorithm $\Xi_\epsilon$.

\begin{algorithm}
\SetAlgoLined
\If{$s=\bot$}{
    \Return $0$;
}
\If{$v[(1-i)]=2$}{
    \Return $3s+i+(-1)^s(2i-1)$
}\If{$v[(1-i)]=\bot$}{
    \Return $3s + i$
}\If{$v[(1-i)]=1$}{
    \Return $3s + i +(-1)^s(1-2i)$
}
\caption{$next\_state_i(v,s)$ function. Code for process $p_i \in \Pi$}\label{alg:nextState}
\end{algorithm}

\begin{theorem}
    The approximate agreement task with precision $\frac{1}{\epsilon}$ is solvable $\forall \epsilon > 0$ by a protocol $A \in \iter$ using any communication primitive $\mathsf{A} \in \smset$.
    Moreover, the required number of rounds $r$ is log-proportional to the precision $\epsilon$: $r \geq log_3(\epsilon)$.
\end{theorem}
\begin{proof}
    We need to prove that $\Xi_\epsilon(\mathcal{I})$ yields a protocol complex consisting of a path of $\epsilon$ edges, as in figure~\ref{fig:b2l3}, made out of vertices $(s,p_{\{0,1\}})$ where $s$ is the internal process state. In $\Xi_\epsilon(\mathcal{I})$, any two adjacent vertices $(s,p_0)$ and $(s',p_1)$ we have that $s-s'\leq 1$. Without loss of generality, we orient the protocol complex such that the leftmost vertex is $(0,p_0)$ and the rightmost $(\frac{3^r-1}{2},p_1)$. 

    First, observe that if the protocol map $\Xi_\epsilon(\mathcal{I})$ behaves as described, then the $\omega$ function is correct: Take any pair of adjacent vertices $(s,p_0)$ and $(s',p_1)$, we have that $|\delta_0(s)-\delta_1(s')|=|\frac{2(s-s')-1}{\epsilon}|\leq \frac{1}{\epsilon}$. Moreover, $\delta_0(0)=0$ and $\delta_1(\frac{3^r-1}{2})=\frac{3^r}{\epsilon}=1$. From here it follows that to have $\frac{1}{\epsilon}$ precision, we need at least that $r\geq log_3(\epsilon)$.

    We proceed by induction on $r$ that $\omega$ and $next\_state_i$ yield the desired protocol complex. The base case, $r=0$ is immediate: $\Xi_\epsilon^0(\mathcal{I})=\mathcal{I}$ where $\mathcal{I}$ is a path of length 1, and $\delta_0(0)=0$, $\delta_1(0)=1$. For the inductive step, because $s-s'\leq 1$, we distinguish two possible process configurations: $\{(s,p_0),(s,p_1)\}$ and $\{(s+1,p_0),(s,p_1)\}$. Assume $s\geq0$ even, thus $\omega(s)=1$.

    Consider the edge $\{(s,p_0),(s,p_1)\}$. In the next iteration, there are 3 possibilities: Both processes read $2$, $p_0$ reads $\bot$ but $p_1$ reads $2$, and $p_1$ reads $\bot$ but $p_0$ reads $2$. The states in the next iteration will be $next\_state_0(1,s)=3s+1$, $next\_state_1(1,s)=3s$, $next\_state_0(\bot,s)=3s$ and $next\_state_1(\bot,s)=3s+1$. Thus, from $\{(s,p_0),(s,p_1)\}$ the next iteration will yield a path of 3 edges: $(3s,p_0)-(3s,p_1)-(3s+1,p_0)-(3s+1,p_1)$. We have that the states of $p_0$ are equal or at most $1$ value bigger than the states of $p_1$. The reasoning is analogous if $s$ is odd.

    Thus, each edge of the $r-1$ iteration protocol complex yields 3 new edges in the next iteration, where the initial vertices are relabeled by the function $next\_state_i(s,\bot)=3s+i$. As a result, $\Xi_\epsilon^r$ will be a path of length $3^r$ such that leftmost vertex is $(0,p_0)$ and the rightmost $(\frac{3^r-1}{2},p_1)$.
\end{proof}

It is remarkable that we can have arbitrarily high $\epsilon$ precision using constant bit complexity. 
    Furthermore, we need $\log_3(\epsilon)$ layers to reach the desired precision. It follows that $\epsilon$-approximate agreement is achievable in $O(log(\epsilon))$ rounds, each using $2$-bit communication primitive entries.

The bounded protocol introduced here resembles the approximate agreement algorithm proposed in~\cite{2process_complexity} for wait-free dynamic networks, where 1-bit messages are used.
We use two bits in our model as we consider that one of the possible bit encodings is used to represent the initial value of the communication primitive. 
Hence, the encoding of three different values is required: those written by the processes ($1$ and $2$), along with an initial value, denoted here as $\bot$, representing an uninitialized state of the communication primitive.

Furthermore, the colored approximate agreement task's output complex is isomorphic to $\lceil \log_3 (\epsilon) \rceil$ iterations of the chromatic subdivision.
Therefore, the protocol map $\Xi_\epsilon$ is equivalent, up to isomorphism, to the full-information protocol but with a constant 2-bit complexity per iteration.
In conclusion, for 2 processes to replicate the \textit{full-information} IIS protocol the bit complexity is 2 ($\Theta(1)$). 

The difference between simulating full-information for the two-process case and the case $n+1 \ge 3$ is fundamentally topological. 
When the input complex is one-dimensional (i.e., two processes), chromatic subdivision preserves a constant bound
on the degree of vertices.
Consequently, both the maximum degree of the protocol complex and the size of cliques in the corresponding indistinguishability graphs remain constant.
This bounded structure allows the protocol to operate with a constant number of encodings.
In higher dimensions ($n \ge 3$), however, each application of the chromatic subdivision introduces vertices of increasing degree, and the corresponding indistinguishability graphs grow accordingly with the number of iterations of the full-information protocol.
As a result, constant encodings are no longer sufficient, as shown in the previous sections.

\section{Discussion and Open Questions}
\label{sec:end}
This work studies the simulation of \textit{full-information} protocols using bounded iterated memory. 
Using tools from combinatorial topology, we derive asymptotic lower and upper bounds for simulating iterated full-information protocols, which apply to several iterated shared-memory models studied in the literature, including collect, atomic snapshot, and immediate snapshot.

These results extend recent work on bounded-memory computability
\cite{delporte2024computational}, providing new tight bounds on the cost of simulating full-information protocols with bounded bit complexity.
Our results underscore the application of combinatorial topology for distributed systems by identifying necessary and sufficient properties that the protocol complex must satisfy in order to be equivalent to the full-information protocol complex.
Moreover, the Greedy Star algorithm is the first round-efficient construction for simulating unbounded full-information protocols using a bounded number of bits per round.
This contribution highlights the power of combinatorial topology not only as an analytical tool for deriving lower bounds in distributed systems, but also as a constructive tool for designing algorithms that exploit the topological structure of computation.

While full-information protocols are a powerful abstraction for reasoning about distributed computability, they are prohibitively expensive to implement in practice.
For a fixed number of bits $b$, the round complexity of iterated full-information protocols grows superexponentially with the number of processes and iterations.
Mitigating this growth requires increasing the bit complexity per process, which scales as $\Theta(r n \log n)$. 
This is substantially more efficient than the unbounded full-information protocol (Algorithm~\ref{alg:FIP}), reducing the bit complexity from $\Theta(n^{r-1})$ to $\Theta(r n \log n)$ bits.
However, since real systems have bounded bit complexity, $b$ cannot be made arbitrarily large, and even with a modest number of processes, only a small number of full-information iterations can be simulated.

Several research directions remain open.
A natural direction is to extend the techniques developed here to tasks beyond simplex agreement on a full-information protocol complex, such as renaming and weak symmetry breaking~\cite{distCompTopo}.
Another direction is to study bounded implementations under weaker or stronger communication primitives within $\iter$.
This requires analyzing the protocol complexes induced by such primitives and deriving corresponding $f$-vector bounds. 
Stronger primitives may allow simulations with improved asymptotic complexity, while weaker primitives may require additional rounds or larger encodings.
Moreover, one can consider this question in the context of bounded-memory \emph{adversarial} models~\cite{HR10,KRH18}. 
Another promising direction is to study randomized algorithms in this framework. Recent work~\cite{Fraign21} introduces probabilistic protocol complexes, where simplices represent executions with associated probabilities. Such models could yield improved bounds by allowing protocols to disregard low-probability executions.

Finally, an open question is whether the round complexity upper bound for bounded snapshot-based implementations can be reduced.
Specifically, given $b$-bit snapshot entries per process per round, it remains to determine whether an unbounded full-information snapshot protocol can be implemented using fewer than $O\bigl((n!)^{r-1} \cdot 2^{n - b} \cdot n\bigr)$ rounds.

\bibliographystyle{cas-model2-names}
\bibliography{refs}

@inproceedings{prequel1,
  title={Space-Time Trade-Off in Bounded Iterated Memory},
  author={Toyos-Marfurt, Guillermo and Kuznetsov, Petr},
  booktitle={International Symposium on Stabilizing, Safety, and Security of Distributed Systems},
  pages={458--474},
  year={2025},
  organization={Springer}
}

@InProceedings{prequel2,
author="Toyos-Marfurt, Guillermo
and Kuznetsov, Petr",
title="On the Bit Complexity of Iterated Memory",
booktitle="Structural Information and Communication Complexity",
year="2024",
pages="456--477",
organization={Springer},
isbn="978-3-031-60603-8"
}

@book{AttiyaWelchBook,
  author       = {Hagit Attiya and
                  Jennifer L. Welch},
  title        = {Distributed computing - fundamentals, simulations, and advanced topics
                  {(2.} ed.)},
  series       = {Wiley series on parallel and distributed computing},
  publisher    = {Wiley},
  year         = {2004},
  isbn         = {978-0-471-45324-6}
}

@article{AfekSnapshots93,
author = {Afek, Yehuda and Attiya, Hagit and Dolev, Danny and Gafni, Eli and Merritt, Michael and Shavit, Nir},
title = {Atomic snapshots of shared memory},
year = {1993},
issue_date = {Sept. 1993},
publisher = {Association for Computing Machinery},
address = {New York, NY, USA},
volume = {40},
number = {4},
issn = {0004-5411},
journal = {J. ACM},
month = sep,
pages = {873–890},
numpages = {18},
}

@inproceedings{ImmediateSnap,
author = {Borowsky, Elizabeth and Gafni, Eli},
title = {Immediate atomic snapshots and fast renaming},
year = {1993},
isbn = {0897916131},
publisher = {Association for Computing Machinery},
address = {New York, NY, USA},
booktitle = {Proceedings of the Twelfth Annual ACM Symposium on Principles of Distributed Computing},
pages = {41–51},
numpages = {11},
location = {Ithaca, New York, USA},
series = {PODC '93}
}

@article{chromsubdivision,
author = {Dmitry N. Kozlov},
title = {{Chromatic subdivision of a simplicial complex}},
volume = {14},
journal = {Homology, Homotopy and Applications},
number = {2},
publisher = {International Press of Boston},
pages = {197 -- 209},
year = {2012},
}

@Book{distCompTopo,
  author = 	 {Maurice Herlihy and Dmitry N. Kozlov and Sergio Rajsbaum},
  ALTeditor = 	 {},
  title = 	 {Distributed Computing Through Combinatorial Topology},
  publisher = 	 {Morgan Kaufmann},
  year = 	 {2014},
  OPTkey = 	 {},
  OPTvolume = 	 {},
  OPTnumber = 	 {},
  OPTseries = 	 {},
  OPTaddress = 	 {},
  OPTedition = 	 {},
  OPTmonth = 	 {},
  OPTnote = 	 {},
  OPTannote = 	 {}
}

@article{brooks_1941, title={On colouring the nodes of a network}, volume={37}, number={2}, journal={Mathematical Proceedings of the Cambridge Philosophical Society}, publisher={Cambridge University Press}, author={Brooks, R. L.}, year={1941}, pages={194–197}}

@Article{Karp-NP,
  author = 	 {Richard M. Karp},
  title = 	 {Reducibility Among Combinatorial Problems},
  journal = 	 {Complexity of Computer Computations},
  year = 	 {1972},
  pages = 	 {85-103}
}

@article{fubiniNumbers,
 ISSN = {00029890, 19300972},
 author = {O. A. Gross},
 journal = {The American Mathematical Monthly},
 number = {1},
 pages = {4--8},
 publisher = {Mathematical Association of America},
 title = {Preferential Arrangements},
 volume = {69},
 year = {1962}
}

@inproceedings{delporte2024computational,
author = {Delporte-Gallet, Carole and Fauconnier, Hugues and Fraigniaud, Pierre and Rajsbaum, Sergio and Travers, Corentin},
title = {The Computational Power of Distributed Shared-Memory Models with Bounded-Size Registers},
  booktitle =     {PODC},
year = {2024},
isbn = {9798400706684},
pages = {310–320},
location = {Nantes, France},
  publisher =     {Association for Computing Machinery},
  address =     {New York, NY, USA},
}

@article{aproxag,
author = {Dolev, Danny and Lynch, Nancy A. and Pinter, Shlomit S. and Stark, Eugene W. and Weihl, William E.},
title = {Reaching Approximate Agreement in the Presence of Faults},
year = {1986},
issue_date = {July 1986},
publisher = {Association for Computing Machinery},
journal = {Journal of the ACM},
address = {New York, NY, USA},
volume = {33},
number = {3},
issn = {0004-5411},
}

@article{consensus,
author = {Fischer, Michael J. and Lynch, Nancy A. and Paterson, Michael S.},
title = {Impossibility of Distributed Consensus with One Faulty Process},
year = {1985},
issue_date = {April 1985},
publisher = {Association for Computing Machinery},
address = {New York, NY, USA},
volume = {32},
number = {2},
issn = {0004-5411},
journal = {J. ACM},
month = {apr},
pages = {374–382},
numpages = {9}
}

@InProceedings{2process_complexity,
author="Delporte-Gallet, Carole
and Fauconnier, Hugues
and Rajsbaum, Sergio",
editor="Richa, Andrea Werneck
and Scheideler, Christian",
title="Communication Complexity of Wait-Free Computability in Dynamic Networks",
booktitle="Structural Information and Communication Complexity",
year="2020",
publisher="Springer International Publishing",
address="Cham",
pages="291--309",
isbn="978-3-030-54921-3"
}

@Inbook{fvector,
author="Ziegler, G{\"u}nter M.",
title="Shellability and the Upper Bound Theorem",
bookTitle="Lectures on Polytopes: Updated Seventh Printing of the First Edition",
year="1995",
publisher="Springer New York",
pages="231--290"
}

@article{HS99,
 Author = {Maurice Herlihy and Nir Shavit},
 Title = {The topological structure of asynchronous computability},
 Journal = jacm,
 Volume = 46,
 number = 2,
 Year = 1999,
 pages = "858--923",
 }

@inproceedings{GKM14-podc,
  author    = {Eli Gafni and
               Petr Kuznetsov and
               Ciprian Manolescu},
  title     = {A generalized asynchronous computability theorem},
  booktitle = {PODC},
  year      = {2014}
}

@inproceedings{KRH18,
 author = {Petr Kuznetsov and Thibault Rieutord and Yuan He},
 title = {An Asynchronous Computability Theorem for Fair Adversaries},
 booktitle = {Proceedings of the ACM Symposium on Principles of Distributed Computing},
 series = {PODC 2018},
 year = {2018},
 pages = {1--10}
}

@article{mosesCons,
author = {Moses, Yoram and Rajsbaum, Sergio},
title = {A Layered Analysis of Consensus},
journal = {SIAM Journal on Computing},
volume = {31},
number = {4},
pages = {989-1021},
year = {2002}
}

@inproceedings{HerSAS90,
author = {Herlihy, Maurice and Rajsbaum, Sergio and Tuttle, Mark R.},
title = {Unifying synchronous and asynchronous message-passing models},
year = {1998},
isbn = {0897919777},
publisher = {Association for Computing Machinery},
address = {New York, NY, USA},
booktitle = {Proceedings of the Seventeenth Annual ACM Symposium on Principles of Distributed Computing},
pages = {133–142},
numpages = {10},
location = {Puerto Vallarta, Mexico},
series = {PODC '98}
}

@inproceedings{Raj10Shared,
author = {Rajsbaum, Sergio},
title = {Iterated shared memory models},
year = {2010},
isbn = {3642121993},
publisher = {Springer-Verlag},
address = {Berlin, Heidelberg},
booktitle = {Proceedings of the 9th Latin American Conference on Theoretical Informatics},
pages = {407–416},
numpages = {10},
location = {Oaxaca, Mexico},
series = {LATIN'10}
}

@article{ELRAD1982155,
title = {Decomposition of distributed programs into communication-closed layers},
journal = {Science of Computer Programming},
volume = {2},
number = {3},
pages = {155-173},
year = {1982},
issn = {0167-6423},
author = {Tzilla Elrad and Nissim Francez}
}

@article{HERLIHY20133,
title = {Power and limits of distributed computing shared memory models},
journal = {Theoretical Computer Science},
volume = {509},
pages = {3-24},
year = {2013},
note = {Structural Information and Communication Complexity},
issn = {0304-3975},
author = {Maurice Herlihy and Sergio Rajsbaum and Michel Raynal},
}

@article{winkler2024time,
  title={The time complexity of consensus under oblivious message adversaries},
  author={Winkler, Kyrill and Paz, Ami and Galeana, Hugo Rincon and Schmid, Stefan and Schmid, Ulrich},
  journal={Algorithmica},
  volume={86},
  number={6},
  pages={1830--1861},
  year={2024},
  publisher={Springer}
}

@inproceedings{nishi,
  author       = {Susumu Nishimura},
  title        = {Schlegel Diagram and Optimizable Immediate Snapshot Protocol},
  booktitle    = {21st International Conference on Principles of Distributed Systems,
                  {OPODIS} 2017},
  pages        = {22:1--22:16},
  year         = {2017},
}

@inproceedings{BG93a,
  author =     {Elizabeth Borowsky and Eli Gafni},
  title =     {Immediate atomic snapshots and fast renaming},
  booktitle =     {PODC},
  year =     {1993},
  isbn =     {0-89791-613-1},
  pages =     {41--51},
  location =     {Ithaca, New York, United States},
  doi =         {http://doi.acm.org/10.1145/164051.164056},
  publisher =     {ACM Press},
  address =     {New York, NY, USA},
}

@inproceedings{SZ93,
  booktitle = proc # " 25th " # stoc,
  month = may,
  Year = 1993,
  confaddress = "San Diego, California",
  confeditor = "Alok Aggrawal",
  Author = {Saks, Michael and Fotios Zaharoglou},
  Title = "Wait-Free $k$-Set Agreement is Impossible: The Topology of
Public Knowledge",
  pages = "101--110"
  }

@inproceedings{GR10-opodis,
  author    = {Eli Gafni and
               Sergio Rajsbaum},
  title     = {Distributed Programming with Tasks},
  booktitle = {Principles of Distributed Systems - 14th International Conference,
               {OPODIS} 2010},
  pages     = {205--218},
  year      = {2010}
}

@inproceedings{BG97,
  author =     {Elizabeth Borowsky and Eli Gafni},
  title =     {A simple algorithmically reasoned characterization of
                  wait-free computation (extended abstract)},
  booktitle =     {PODC '97: Proceedings of the sixteenth annual ACM
symposium
                  on Principles of distributed computing},
  year =     {1997},
  isbn =     {0-89791-952-1},
  pages =     {189--198},
  location =     {Santa Barbara, California, United States},
  doi =         {10.1145/259380.259439},
  publisher =     {ACM Press},
  address =     {New York, NY, USA},

}

@inproceedings{HR10,
  author    = {Maurice Herlihy and
               Sergio Rajsbaum},
  title     = {The topology of shared-memory adversaries},
  booktitle = {PODC},
  year      = {2010},
  pages     = {105-113}
}

@article{hoest,
title = "Toward a topological characterization of asynchronous complexity",
author = "Gunnar Hoest and Nir Shavit",
year = "2006",
volume = "36",
pages = "457--497",
journal = "SIAM Journal on Computing",
issn = "0097-5397",
publisher = "Society for Industrial and Applied Mathematics (SIAM)",
number = "2"
}

@InProceedings{bouzidStrg,
author="Bouzid, Zohir
and Gafni, Eli
and Kuznetsov, Petr",
editor="Aguilera, Marcos K.
and Querzoni, Leonardo
and Shapiro, Marc",
title="Strong Equivalence Relations for Iterated Models",
booktitle="Principles of Distributed Systems",
year="2014",
publisher="Springer International Publishing",
address="Cham",
pages="139--154",
}

@misc{gob26,
      title={Distributed Knowledge in Simplicial Models}, 
      author={Éric Goubault and Jérémy Ledent and Sergio Rajsbaum},
      year={2026},
      eprint={2602.06945},
      archivePrefix={arXiv},
      primaryClass={cs.LO},
      url={https://arxiv.org/abs/2602.06945}, 
}

@article{gob21,
title = {A simplicial complex model for dynamic epistemic logic to study distributed task computability},
journal = {Information and Computation},
volume = {278},
pages = {104597},
year = {2021},
note = {Special Issue on Ninth International Symposium on Games, Automata, Logics and Formal Verification (GandALF 2018)},
issn = {0890-5401},
author = {Éric Goubault and Jérémy Ledent and Sergio Rajsbaum},
}

@article{HW90,
author="Maurice Herlihy and  Jeannette M. Wing",
title="Linearizability: A Correctness Condition for Concurrent Objects",
journal=" ACM Trans. Program. Lang. Syst.",
volume= 12,
number=3,
pages="463--492",
year= 1990
}

@inproceedings{Fraign21,
author = {Fraigniaud, Pierre and Gelles, Ran and Lotker, Zvi},
title = {The Topology of Randomized Symmetry-Breaking Distributed Computing},
year = {2021},
isbn = {9781450385480},
publisher = {Association for Computing Machinery},
address = {New York, NY, USA},
booktitle = {Proceedings of the 2021 ACM Symposium on Principles of Distributed Computing},
pages = {415–425},
numpages = {11},
location = {Virtual Event, Italy},
series = {PODC'21}
}

@article{attiyaCollect,
  author       = {Hagit Attiya and
                  Arie Fouren and
                  Eli Gafni},
  title        = {An adaptive collect algorithm with applications},
  journal      = {Distributed Comput.},
  volume       = {15},
  number       = {2},
  pages        = {87--96},
  year         = {2002},
}

\end{document}